\documentclass[letter]{aa}
\usepackage{color} 
\usepackage{graphicx}
\usepackage{amsmath,epsfig,rotating,amssymb}
\usepackage{natbib}
\def\simless{\mathbin{\lower 3pt\hbox
   {$\rlap{\raise 5pt\hbox{$\char'074$}}\mathchar"7218$}}}
\def\simgreat{\mathbin{\lower 3pt\hbox
   {$\rlap{\raise 5pt\hbox{$\char'076$}}\mathchar"7218$}}}

\bibpunct{(}{)}{;}{a}{}{,} 

\begin{document}

\author{R. S. Klessen\inst{1,2}, P. Hennebelle\inst{2}}

\institute{Zentrum f\"{u}r Astronomie der Universit\"{a}t Heidelberg, Institut 
f\"{u}r Theoretische Astrophysik, 69120 Heidelberg, Germany 
\and 
Laboratoire de radioastronomie, UMR 8112 du CNRS, 
{\'E}cole normale sup{\'e}rieure et Observatoire de Paris,
 24 rue Lhomond, 75231 Paris cedex 05, France 
}

\title{Accretion-Driven Turbulence as Universal Process: Galaxies, Molecular Clouds, and Protostellar Disks }

\authorrunning{Klessen \& Hennebelle}

\titlerunning{Accretion-Driven Turbulence}

\abstract
{ Complex turbulent motions are ubiquitously observed in many astrophysical systems.  
The origin of this turbulence, however, is still poorly understood. 
%
}
{ When cosmic structures form, they grow in mass via accretion from their surrounding environment.  We propose that this accretion is able to drive internal turbulent motions in a wide range of astrophysical objects and study this process in the case of galaxies, molecular clouds and protoplanetary disks. 
}
{We use a combination of numerical simulations and analytical arguments to predict the  
level of turbulence as a function of the accretion rate, the dissipation scale, and the density 
contrast, and compare with observational data. 
}
{{ We find that in Milky Way type galaxies the observed level of turbulence in the interstellar medium can be explained by accretion, provided that the galaxies gain mass at a rate comparable to the rate at which they form stars. This process is particularly relevant in the extended outer disks beyond the star-forming radius. 
In order to drive turbulence in  dwarf galaxies,  the accretion rate needs to exceed the star formation rate by a large factor and we  expect other sources to dominate.}
We also calculate the rate at which molecular clouds grow in mass when they build up from the atomic component of the galactic gas and find that their internal turbulence is likely to be driven by accretion as well. It is the very process of cloud formation that excites turbulent motions on small scales by establishing the turbulent cascade. 
In the case of T~Tauri disks, we show that accretion can drive subsonic turbulence  at the observed level if the rate at which gas falls onto the disk is comparable to the rate at which disk material accretes onto the central star. This also explains the observed relation of accretion rate and stellar mass, $\dot{M} \propto M_\star^{1.8}$. 
The efficiency required to convert infall motion into turbulence is of the order of a few percent in all three cases.
}
{We conclude that accretion-driven turbulence is a universal concept with far-reaching implications for a wide range of astrophysical objects.}

\maketitle

\section{Introduction}
\label{sec:intro}

Astrophysical fluids on virtually all scales are characterized by highly complex turbulent motions. This ranges from the gas between galaxies to the interstellar medium (ISM) within them, as well as from individual star-forming molecular clouds down to the protostellar accretions disks that naturally accompany stellar birth, and has far reaching consequences for cosmic structure formation. For example, it is the complex interplay between supersonic turbulence in the ISM and self-gravity in concert with magnetic fields, radiation, and thermal pressure that determines when and where stars form in the Galaxy \citep{MacLow:2004p2713,Larson:2005p657,BallesterosParedes:2007p6136,Mckee:2007p34}. 
Similar is true for protostellar accretion disks, where turbulent motions cause angular momentum redistribution and thus determine the rate at which material accretes onto the central star and the likelihood to build up planets and planetary systems.

Yet, despite its ubiquity and importance, very little is known about the origin of astrophysical turbulence. The number of possible sources is large and varies strongly depending on the physical scale under consideration. For a discussion of possible sources of ISM turbulence, see e.g. \citet{MacLow:2004p2713} or \citet{Elmegreen:2004p6143}. Here we attempt to argue that it is the accretion process, that inevitably goes along with any astrophysical structure formation, let it be the birth of galaxies or stars, that drives the observed turbulent motions. We propose that this process is universal and makes significant contributions to the turbulent energy on all scales \cite[see also][]{Field:2008p17384}.
%
We ask: Does the accretion flow onto galaxies, onto dense clouds in the ISM within these galaxies, and finally onto the protostellar accretion disks that accompany stellar birth within these clouds provide enough energy to account for the observed internal motions? What is the expected efficiency for the conversion of kinetic energy associated with the infalling material into kinetic energy associated with internal turbulence? Our analysis leads us to believe that accretion is indeed an important driver of turbulence on all scales observed. 

We  structure our discussion as follows: We introduce the concept of accretion-driven turbulence in Section \ref{sec:concept}. We estimate the energy input associated with accretion and compare it to the energy needed to compensate for the decay of turbulent motions assuming overall steady state. Under typical conditions the energy gain from accretion exceeds the energy loss by the decay of turbulence by far. However, we do not know the efficiency with which infall motions are converted into random turbulent motions. To get a handle on this quantity, we resort to numerical simulations of convergent astrophysical flows for guidance and propose a theoretical explanation of the trends inferred from the simulations. 
In Section \ref{sec:galaxies} we apply our method to galactic scales and propose that the turbulent velocity dispersion measured in the disk of the Milky Way and other galaxies are  caused by accretion streams that originate in or pass through the halo. We then turn to the scales of individual interstellar gas clouds and argue in Section \ref{sec:clouds} that it is the process of cloud formation that drives their internal turbulent motions. Our third applications lies on even smaller scales. In Section \ref{sec:disks} we speculate about the origin of turbulence in accretion disks. We focus our discussion on protostellar accretion disks { during the late stages of the evolution (class 2 and 3 phases),} but we note that similar arguments may apply to the accretion disks around black holes in active galactic nuclei. Finally, we conclude in Section \ref{sec:end}.

\section{Basic Concept} 
\label{sec:concept}

\subsection{Energy Balance}
\label{subsec:decay-rate}
Several numerical studies \citep{MacLow:1998p384,Stone:1998p12501,Padoan:1999p516,MacLow:1999p251,Elmegreen:2000p9513} have demonstrated that supersonic turbulence decays on a timescale that is equivalent to the turbulent crossing time,
\begin{eqnarray}
\tau_{\rm d} &\approx&  \frac{L_{\rm d}}{\sigma}\,,
\end{eqnarray}
where $L_{\rm d}$ is the driving scale and $\sigma$ is the 3-dimensional velocity dispersion. This holds regardless whether the gas is magnetized or not and also extends into the subsonic regime. The exact value of $L_{\rm d}$ is not well constrained by the observational data {  and needs to be chosen with care for each system under consideration. We note, however, that it is a universal feature of the objects we study here that the bulk of the kinetic energy is carried by the largest spatial modes, consistent with turbulence being driven} from the outside \cite[see, e.g.][for nearby molecular clouds]{Ossenkopf:2002p334,Brunt:2003p17014,Brunt:2009p16966}. 

 The total loss of turbulent kinetic energy, $E = 1/2 M \sigma^2$, to a system with total mass $M$
 through turbulent decay sums up to  
\begin{eqnarray} 
\label{eqn:dissip}
\dot{E}_{\rm decay} & \approx&  \frac{E}{\tau_{\rm d}} = - \frac{1}{2} \frac{M \sigma^3}{L_{\rm d}}  
\end{eqnarray}

When the system accumulates mass at a rate $\dot{M}$ the associated kinetic energy is 
\begin{eqnarray} 
\label{eqn:dissip1}
\dot{E}_{\rm in}&=& \frac{1}{2} \dot{M}_{\rm in} v^2_{\rm in}\, 
\end{eqnarray}
where $v_{\rm in}$ is the infall velocity.

We introduce an efficiency factor 
\begin{eqnarray} 
\label{eqn:dissip2}
\epsilon = \left|\frac{\dot{E}_{\rm decay}}{\dot{E}_{\rm in}}\right|
\end{eqnarray}
which represents the fraction, $\epsilon$, of the available accretion energy required to sustain the observed 
turbulent velocities. 

For the hypothesis of accretion-driven turbulence to work, clearly  ${\dot{E}_{\rm in}} \ge {\dot{E}_{\rm decay}}$
is required. This is usually true as we discuss in the Sections below.  We note, however, that the fraction of
the infall energy that actually is converted into random turbulent motions is very difficult to estimate.
{ Clearly some fraction of the accretion energy turns into heat and is radiated away. In addition, if the system is highly inhomogeneous with most of the  mass residing in high-densities clumps with low volume filling factor, most of the incoming flux will feed the tenuous interclump medium rather than the dense clumps, and again, not contribute directly to driving their internal turbulence. }
This is taken into account in the efficiency factor.
 Numerical experiments indicate that $\epsilon$ depends on the density contrast between the infalling gas and the material in the system under consideration (see Section \ref{subsec:input-rate}). For molecular clouds forming in convergent flows $\epsilon$ is of order of 0.01 to 0.1. If these values are representative for other systems, then in general $\dot{E}_{\rm in}$
needs to be 10 to 100 times larger than $\dot{E}_{\rm decay}$.

\subsection{Estimate of Efficiency}
\label{subsec:input-rate}

To estimate the efficiency at which accretion energy is converted into turbulent energy we resort to numerical simulations of converging flows { \cite[e.g.][]{Audit:2005p194,Heitsch:2005p12601,Heitsch:2006p8314,Folini:2006p12636,VazquezSemadeni:2006p13319,VazquezSemadeni:2007p47,Hennebelle:2008p832,Audit:2010p26999,Banerjee:2009p6188}.} These simulations consider two colliding flows of diffuse gas which produce strong density fluctuations of cold gas. { The incoming velocity is initially supersonic with respect to the cold and dense gas which forms under the influence of cooling and ram pressure.
It is generally found, that the resulting turbulence in this component is comparable to the observed values in Galactic molecular clouds.}

\begin{figure}[htbp]
\centerline{\psfig{file=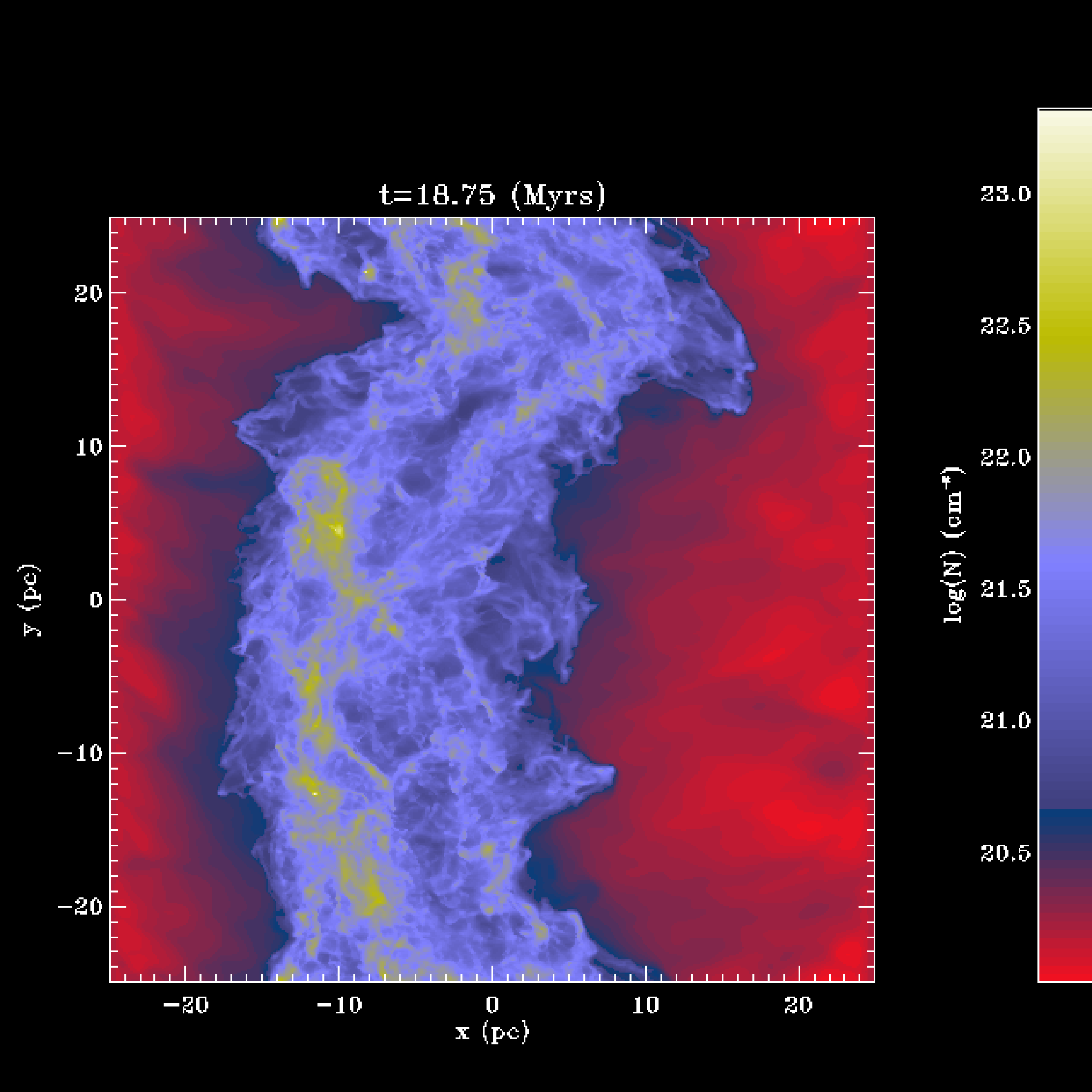,width=8.5cm}}
\caption{Column density at $t=18.75\,$Myr in the high resolution colliding flow calculation. }
\label{col_dens}
\end{figure}

The simulations reported here are very similar to those presented by \citet{Hennebelle:2008p832} and \citet{Audit:2010p26999}. 
They have been performed  with the adaptive mesh-refinement magnetohydrodynamics  code RAMSES \citep{Teyssier:2002p12758, Fromang:2006p12761} and  include 
magnetic fields (initially uniform and equal to 5$\,\mu$G) and self-gravity. They are either isothermal at $T=50\,$K or start 
with a warm neutral medium at $T=8000\ $K and self-consistently treat cooling processes assuming a { standard  2-phase ISM cooling function.} The gas is injected from the boundary with a density equal to 
1$\,$cm$^{-3}$ and a mean velocity of either 15 or 20$\,$km$\,$s$^{-1}$ on top of which {  fluctuations with an amplitude of 50\%} have been superimposed.
In order to quantify the impact of the numerical resolution (a crucial issue) as well as the influence of the thermal 
structure of the flow, we present the results of four calculations. First, two lower resolution 
calculations with an incoming velocity of 15$\,$km$\,$s$^{-1}$, one isothermal and one in which cooling is treated.
Both  have an initial grid of 256$^3$ computing cells and two further AMR levels are used when the density reaches
a threshold of 80 and 160 cm$^{-3}$.  Second, we also present two simulations with a higher incoming velocity of 20 km s$^{-1}$. One has 
a high resolution and starts with  512$^3$ computing cells and 
four further AMR levels are used when density reaches respectively 50, 100, 400 and 1600$\,$cm$^{-3}$. 
In this  calculation the number of cells is about $5 \times 10^8$. The other has the same resolution as the two lower resolution simulations.
Figure~\ref{col_dens} shows the column density in the computational box for the high resolution run. 

\begin{figure}[htbp]
\centerline{\psfig{file=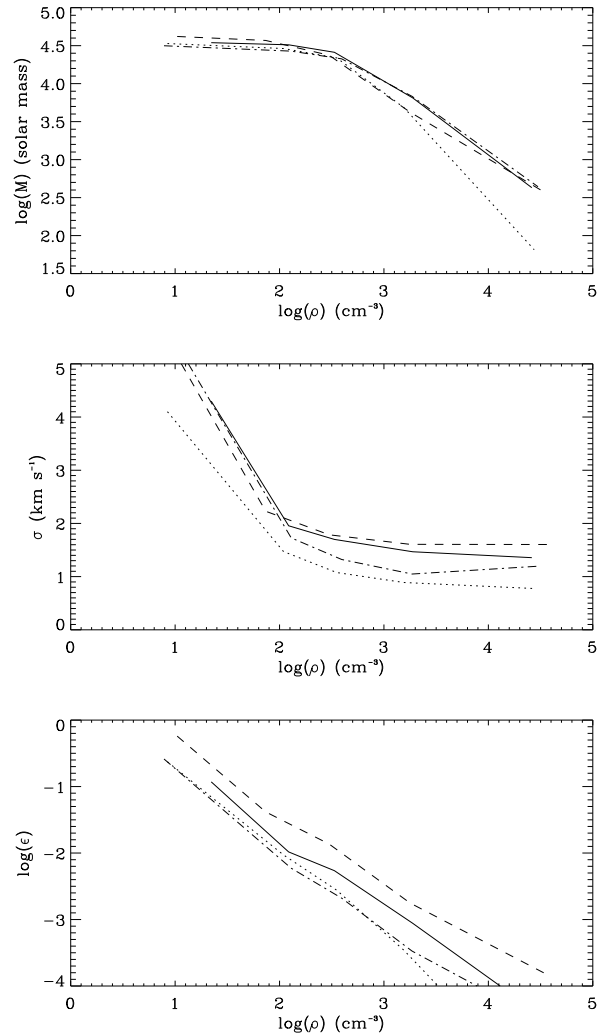,width=8.5cm}}
\caption{Mass, velocity dispersion and efficiency of the energy injection as a function of gas density in four colliding flow calculations.
Solid, dotted and dashed lines show the  cases with standard ISM cooling. The solid one corresponds to the highest numerical resolution and an incoming velocity of 20$\,$km$\,$s$^{-1}$, the dash-dotted line is identical except that it  has a lower resolution, and the dotted line is for a lower resolution simulation and an incming velocity of 15$\,$km$\,$s$^{-1}$. The { dashed-dotted} line corresponds to the purely isothermal calculation (with gas temperature of about 50$\,$K). It has the same resolution than the lower resolution simulations with cooling.}
\label{efficiency}
\end{figure}

Figure~\ref{efficiency} shows the efficiency $\epsilon$ as defined by Eq.~(\ref{eqn:dissip2}), as a function of the gas density. 
That is, we select all cells in the computing domain {  with densities above a certain threshold value $\rho_{\rm t}$ and then compute the 
quantity $\int \rho v^3 dV = M_{\rm t} \sigma_{\rm t}^3$, where $\rho$ and $v$ are the density and velocity of the cells, and $M_{\rm t}$ and $\sigma_{\rm t}$ are total mass and velocity dispersion of the gas above $\rho_{\rm t}$. 
 Finally we divide by $L_{box}$ and 
  by $\dot{E}_{in}$ as defined by Eq.~(\ref{eqn:dissip1}).}
Various trends can be inferred from Fig.~\ref{efficiency}. First, for all the simulations we find that the 
efficiency decreases with the gas density as roughly $1/\rho$. The two low resolution simulations with cooling but different 
incoming velocities are very close to each other. The high resolution simulation exhibits a slightly higher efficiency. This is expected as the numerical dissipation is lower. However, it is larger by only a small factor of 1.5 to 2, suggesting that { our result is reasonably well converged}. Finally we see that the efficiency is larger in the isothermal case by a factor of $\sim 3$ compared to the low resolution runs with cooling. 
This indicates that the thermal structure of the flow has a significant impact on its dynamics \cite[see also][]{Audit:2010p26999}.  Because the isothermal run and the simulations with a 2-phase medium cooling function are significantly different, the discrepancy gives us an estimate of  the uncertainty of the efficiency. { Interestingly, Fig.~\ref{efficiency} shows two different regimes. At low densities the total mass above 
the threshold value decreases slowly with density, while the velocity 
dispersion decreases more steeply. At large densities, however,  the mass
decreases rapidly with density while the velocity dispersion is 
nearly constant. This behavior is discussed further in  Appendix \ref{app:M-sigma}.}

{ We conclude} that for astrophysical systems, such as molecular clouds, of mean density $\bar{\rho}$ accreting gas at density $\sim \rho_{\rm in}$, we typically have
\begin{eqnarray}
\label{eqn:density}
\epsilon \approx \rho_{\rm in} / \bar{\rho}\,.
\end{eqnarray}
This relation is expected to be valid within a factor of a few.

\subsection{Possible Explanation for the $\epsilon$ - $\rho^{-1}$ Relation}
\label{subsec:relation}
{ As the $\epsilon$ - $\rho^{-1}$ relation appears to be both important and interesting, we propose a possible theoretical explanation. 
Consider a turbulent flow with wave numbers ranging from $k_{\rm min}$ to $k_{\rm max}$. For incompressible 
fluids, the power spectrum of the velocity field based on dimensional arguments is expected to be 
$E(k) \propto k^{-5/3}$  \citep{Kolmogorov:1941p10448}. The kinetic energy carried by wave numbers $k$ and larger is given by the integral  $\int _{k}  ^{k_{max}} E(k) dk$, with the total kinetic energy being
$\int _{k_{\rm min}}  ^{k_{\rm max}} E(k) dk$. For compressible media, the scaling relation is more complicated as the density dependency needs to be considered \cite[][]{vonWeizsacker:1951p18341}.} Based on the 
 argument that $ (\rho v^3 / l) / \epsilon _e $ is dimensionless with $\epsilon _e$ being the 
energy flux, it has been suggested { 
\citep{Ferrini:1983p18344,Fleck:1983p14278,Fleck:1996p18391,Kritsuk:2007p14285,Schmidt:2008p829}} that the relation
$E(k) \propto k^{-5/3}$ still holds, provided that $E(k)$ is the power spectrum of $\rho^{1/3} v$ instead of $v$.

Density fluctuations in a turbulent flow follow a roughly log-normal behavior. When identifying clumps and cores, e.g.\ defined  as connected groups of cells/pixels above some thresholds, it has been found that their mass spectrum often follows a power law $dN/dM \propto M^{\alpha}$ with a slope of $\alpha \approx -1.7$ (see, e.g. \citealt{Heithausen:1998p14511} for the observations, or \citealt{Klessen:2001p137}, \citealt{BallesterosParedes:2006p6210}, or \citealt{Hennebelle:2007p12522} for numerical simulations, and \citealt{Hennebelle:2008p6441} for analytical arguments).  
In a convergent flow the biggest clumps cannot be much larger than $l_t = (\rho_{\rm t} / \rho_0)^{-1} \times L_0$ where $\rho _0$ and $L_0$ are the typical density and scales of the large scale flow. Thus, we expect that the largest scale at which clumps denser than $\rho _t$ exist, is $l_{\rm t} \propto \rho_{\rm t}^{-1}$. 

The quantity $\rho_{\rm t}^{2/3} \sigma_{\rm t}^2$ integrated over the cells denser than $\rho_{\rm t}$, is thus expected to be of the order of
$\int _{k_{\rm t}} ^{k_{\rm max}} E(k) dk$, where $E$ is the power spectrum  of $\rho^{1/3} v$ and $k_{\rm t} \simeq 2 \pi / l_{\rm t}$.
Thus, it is found that $\langle \rho _{\rm t} ^{2/3} \sigma_{\rm t}^2 \rangle \approx k _{\rm t} ^{-2/3} \propto l _{\rm t} ^{2/3} \propto \rho _{\rm t} ^{-2/3}$.
This leads to  $\langle \rho _{\rm t}^{2/3} \sigma_{\rm t}^2 \rangle ^{3/2}  \approx \langle \rho _{\rm t} \sigma_{\rm t} ^3 \rangle \propto \rho _{\rm t} ^{-1}$ and after 
multiplication by the volume of the cloud, $M_{\rm t} \sigma_{\rm t}^3 \propto \rho _{\rm t}^{-1}$.

{ So far, we have simply shown that the quantity $M_{\rm t} \sigma_{\rm t}^3$ obtained by integration over
 scales smaller than $l_{\rm t}$ is proportional to $l_{\rm t}$ but it could be the case that the dense
parts of the gas, i.e. regions denser than $\rho_{\rm t}$, have a negligible contribution to this 
integral, in particular because of their low filling factor. However two 
arguments are in disfavor of this statement. First, Hennebelle \& Audit (2007)
have calculated the power spectrum of the kinetic energy of the flow, 
$\rho v^2$, while clipping dense structures above various 
threshold values. As can be seen in their Fig.~(14), the energy contained 
in large-scale motions is unchanged when varying the threshold whereas the energy
contained on the small scales (below about one hundredth of the computing 
box length) decreases with increasing density threshold. It is dominated by high-density structures.
The second argument is also inferred from numerical simulations. The 
power spectrum of $v$ has been calculated in several studies \cite[e.g.][]{Klessen:2000p707,Heitsch:2001p12556, Kritsuk:2007p14285,Federrath:2009p14029} and typically it has been found to be  
$P(v) \propto k ^{-1.9}$. Thus, considering only gas at densities close to the 
mean value of the system, $\bar{\rho}$, we infer that 
$\int _{k_{\rm t}} ^{k_{\rm max}} {\bar{\rho}}^{2/3} v^2 dk \propto 
k_{\rm t}^{-0.9} \propto  l_{\rm t}^{0.9}$ while
$\int _{k_{\rm t}} ^{k_{\rm max}} \rho^{2/3} v^2 dk \propto 
k_{\rm t}^{-2/3} \propto  l_{\rm t}^{2/3}$. 
The implication is that 
as $l_t$ decreases, the contribution of the energy contained in scales
smaller than $l_t$ due to the diffuse gas becomes smaller and smaller
 with respect to the energy contained in the dense gas at these scales. }

The relation, $M_{\rm t} \sigma_{\rm t}^3 \propto \rho _{\rm t}^{-1}$, 
is therefore broadly consistent with 
the trend we  measure. The coefficient seems more difficult to
 predict and as our numerical simulations suggest, it may vary from one flow to another. 
 This requires further investigation. 
 { 
 We note in this context  that the situation is strongly reminiscent of purely incompressible turbulence where dissipation occurs in a subset of space,   in filaments with small filling factor but high vorticity. It can be described by a multi-fractal statistical approach \citep{Frisch:1978p26791,Frisch:1995p26782} to take into account their intermittent nature. For a model based on energy dissipation in shock-generated sheets, see \citet{Boldyrev:2002p26983} extending the theory developed by \citet{She:1994p26987}. 
  }

\section{Turbulence in Galactic Disks}
\label{sec:galaxies}

\subsection{General Considerations}
In our first application we investigate the question as to whether accretion from an external gas reservoir could drive the velocity dispersion observed in spiral galaxies.

The Milky Way, as a typical $L_\star$ galaxy, forms new stars at a rate of $\dot{M}_{\rm SF} \sim 2 - 4\,$M$_\odot\,$yr$^{-1}$. Its gas mass out to $25\,$kpc is $\sim 9 \times 10^9\,$M$_\odot\,$ \citep{Naab:2006p2645,Xue:2008p3090}. Assuming a constant star formation rate, the remaining gas should be converted into stars within about $2 - 4$ Gyr. Similar gas depletion timescales of order of a few billion years are reported for many nearby spiral galaxies \citep{Bigiel:2008p2657}. This is much shorter than the ages of these galaxies which is  $\sim 10^{10}\,$yr. If we discard the possibility that we observe them right at the verge of running out of gas, and instead assume they evolve in quasi steady state, then these galaxies need to be supplied with fresh gas at a rate roughly equal to the star formation rate. 

The requirement of a steady accretion flow onto typical disk galaxies is a natural outcome of cosmological structure formation calculations if baryonic physics is considered consistently. \citet{Dekel:2009p1173} and \citet{Ceverino:2009p1271}, for example, argue that massive galaxies are continuously fed by steady, narrow, cold gas streams that penetrate through the accretion shock associated with the dark matter halo down to the central galaxy. Roughly three quarters of all galaxies forming stars at a given rate are fed by smooth streams \cite[see also][]{Agertz:2009p1291}. On large scales, also the fact that the observed amount of atomic gas in the universe appears to be roughly constant since a redshift of $z \approx 3$ although the stellar content continues to increase, suggests that HI is continuously replenished \citep{Hopkins:2008p9891,Prochaska:2009p9829}.  For our Galaxy, further evidence for a ongoing inflow of low-metallicity material  comes from the presence of deuterium at the solar neighborhood \citep{Linsky:2003p3845} as well as in the Galactic Center \citep{Lubowich:2000p3908}. As deuterium is destroyed in stars and as there is no other known source of deuterium in the Milky Way, it must be of cosmological and extragalactic origin \citep{Ostriker:1975p3929,Chiappini:2002p3920}. 

It is attractive to speculate that the population of high-velocity clouds (HVC) observed around the Milky Way is the visible signpost for high-density peaks in this accretion flow. Indeed the inferred HVC infall rates of $0.5 - 5\,$M$_\odot$yr$^{-1}$  \citep{Wakker:1999p3760,Blitz:1999p4002,Braun:2004p4007,Putman:2006p9534}  are in good agreement with the Galactic star formation rate or with chemical enrichment models 
\cite[see, e.g.][and references therein]{Casuso:2004p3934}. An important question in this context is where and in what form the gas reaches the Galaxy. Recent numerical simulations indicate \citep{Heitsch:2009p8345} that small clouds (with masses less then a few $10^4\,$M$_\odot$) { most likely will dissolve, heat up and merge with the hot halo gas, while larger complexes will be able to deliver cold atomic gas even to the inner disk. 
We explore the idea of continuous gas accretion onto galaxies and argue that this process is a key mechanism for driving interstellar turbulence. }

One of the remarkable features of spiral galaxies is the nearly constant velocity dispersion $\sigma$, e.g.\ as measured in HI emission lines, regardless of galaxy mass and type \citep{Dickey:1990p4035,vanZee:1999p4021,Tamburro:2009p3039}. The inferred  values of $\sigma$ typically fall in a range between $10\,$km$\,$s$^{-1}$ and $20\,$km$\,$s$^{-1}$ \citep{Bigiel:2008p2657,Walter:2008p2669} and extend well beyond the optical radius of the galaxy with only moderate fall-off as one goes outwards. It is interesting in this context that the transition from the star-forming parts of the galaxy to the non-star-forming outer disk seems not to cause significant changes in the velocity dispersion \citep{Tamburro:2009p3039}. 
This apparent independence from stellar sources sets severe constraints on the physical processes that can drive the observed level of turbulence. 

Several possibilities have been discussed in the literature \citep{MacLow:2004p2713,Elmegreen:2004p6143,Scalo:2004p6144}. Large-scale gravitational instabilities in the disk, i.e.\ spiral density waves, can potentially provide sufficient energy \cite[e.g.][]{Li:2005p6076}. However, the efficiency of this process and the details of the coupling mechanism are not well understood. 
Similar holds for the magneto-rotational instability (MRI) which has been identified as a main source of turbulence in protostellar accretion disks \citep{Balbus:1998p5199}. Although we have ample evidence of the presence of  large-scale magnetic fields \citep{Heiles:2005p5988,Beck:2007p5939} there is some debate whether the MRI can provide enough energy to explain the observed levels of turbulence \citep{Beck:1996p6046,Sellwood:1999p5200, Dziourkevitch:2004p5202,Piontek:2007p5212}. For the star-forming parts of spiral galaxies, clearly stellar feedback in form of expanding HII bubbles, winds, or supernova explosions plays an important role. \citet{MacLow:2004p2713} show that the energy and momentum input from supernovae is a viable driving mechanism for interstellar turbulence. However, this approach clearly fails in the extended outer HI disks observed around most spiral galaxies and it also fails in low-surface brightness galaxies. Here accretion driven turbulence seems a viable option \cite[see, e.g.][]{Santillan:2007p17065}.


\subsection{Energy Input Rate}

In order to calculate the energy input rate from the accretion of cold gas we need to know the velocity $v_{\rm in}$ with which this gas falls onto the disk of the galaxy and the efficiency $\epsilon$ with which the kinetic energy of the infalling gas is converted into ISM turbulence. As the cold accretion flow originates from the outer reaches of the halo and beyond and because it lies in the nature of these cold streams that gas comes in almost in free fall, $v_{\rm in}$ can in principle be as high as the escape velocity $v_{\rm esc}$ of the halo. For the Milky Way in the solar neighborhood $v_{\rm esc} \sim 550\,$km$\,$s$^{-1}$ \citep{Fich:1991p3643,Smith:2007p3637}. However, numerical experiments indicate that the inflow velocity of cold streams is of order of the virialization velocity of the halo \citep{Dekel:2009p1173} which typically is $\sim 200\,$km$\,$s$^{-1}$. The actual impact velocity with which this gas interacts with disk material will also depend on the sense of rotation. Streams which come in co-rotating with the disk will have smaller impact velocities  than material that comes in counter-rotating. To relate to quantities that are easily observable and to within the limits of our approximations we adopt $v_{\rm in} = v_{\rm rot}$ as our fiducial value, but note that considerable deviations are possible. 
We also note that even gas that shocks at the virial radius and thus heats up to $10^5 - 10^6\,$K, may cool down again and some fraction of it may be available for disk accretion. This gas can condense into higher-density clumps that sink towards the center and replenish the disk  \citep{Peek:2009p9660}. Again, $v_{\rm in} \approx v_{\rm rot}$ is a reasonable estimate.

We can now calculate the energy input rate associated with this accretion flow as 
\begin{eqnarray}
\label{eqn:input-MW}
\dot{E}_{\rm in}&=& \frac{1}{2} \dot{M}_{\rm in} v^2_{\rm in}\nonumber \\
&=& + 1.3 \times 10^{40} \,\mbox{erg}\,\mbox{s}^{-1}\cdot \nonumber \\
&&\cdot \,\left(\frac{\dot{M}_{\rm in}}{1\,\mbox{M}_{\odot}\,\mbox{yr}^{-1}}\right)
\left(\frac{v_{\rm in}}{200\,\mbox{km}\,\mbox{s}^{-1}}\right)^2\,.
\end{eqnarray}
By the same token, the energy loss through the decay of turbulence is 
\begin{eqnarray} 
\label{eqn:dissip-MW}
\dot{E}_{\rm decay} & \approx&  \frac{E}{\tau_{\rm d}} = - \frac{1}{2} \frac{M \sigma^3}{L_{\rm d}}  
\nonumber \\
&\approx& - 3.2 \times 10^{39} \,\mbox{erg}\,\mbox{s}^{-1} \cdot\,  \nonumber \\
&&\cdot\left(\frac{M}{10^{9}\,\mbox{M}_{\odot}}\right)
\left(\frac{\sigma}{10\,\mbox{km}\,\mbox{s}^{-1}}\right)^3 
\left(\frac{L_{\rm d}}{100 \,\mbox{pc}}\right)^{-1}.
\end{eqnarray}
Given an efficiency $\epsilon$ as defined in Eq.~\ref{eqn:dissip2} the mass accretion rate required to compensate for this energy loss is
\begin{eqnarray}
\label{eqn:mdot}
\dot{M}_{\rm in} &=& \frac{1}{\epsilon}\frac{2 \dot{E}_{\rm decay}}{v^2_{\rm in}} = \frac{1}{\epsilon}\frac{M\sigma^3}{L_{\rm d} v^2_{\rm in}}\nonumber \\
&\approx& 0.25\,\mbox{M}_\odot\,\mbox{yr}^{-1} \,\cdot
\frac{1}{\epsilon}
\cdot\left(\frac{M}{10^{9}\,\mbox{M}_{\odot}}\right)
\left(\frac{\sigma}{10\,\mbox{km}\,\mbox{s}^{-1}}\right)^3\nonumber\\
&&\cdot\left(\frac{L_{\rm d}}{100 \,\mbox{pc}}\right)^{-1}
\left(\frac{v_{\rm in}}{200\,\mbox{km}\,\mbox{s}^{-1}}\right)^{-2}\,.
\end{eqnarray}

\subsection{The Milky Way}
\label{subsec:MW}

Current mass models of the Milky Way \citep{Xue:2008p3090} indicate a total mass including dark matter of about $1\times 10^{12}\,$M$_\odot$ out to the virial radius at $\sim 250\,$kpc. The resulting rotation curve is $220\,$km$\,$s$^{-1}$ at the solar radius $R_\odot \approx 8.5\,$kpc and it declines to values slightly below $200\,$km$\,$s$^{-1}$ at a radius of $60\,$kpc. The total mass in the disk in stars and cold gas is estimated to be $\sim 6 \times 10^{10}\,$M$_\odot$. Assuming a global baryon fraction of 17\% this corresponds to 40\% of all the baryonic mass within the virial radius and  implies that roughly the same amount of baryons is in an extended halo in form of hot and tenuous gas.  The gaseous disk of the Milky Way can be decomposed into a number of different phases. We follow \citet{Ferriere:2001p2879} and consider molecular gas (as traced, e.g. via its CO emission) as well as atomic hydrogen gas (as observed, e.g. by its $21\,$cm emission). The adopted values are summarized in Table \ref{tab:MW}. We neglect the hot ionized medium in our analysis, as Galactic HII regions are produced and heated predominantly by the UV radiation from massive stars and thus should not be included here. We note that 95\% of the turbulent kinetic energy is carried by the atomic component. 

\begin{table}[t]
\caption{Properties of gas components of the Milky Way. }
{\begin{center}
\begin{tabular}{rcc}
\hline
Component  &molecular gas & atomic gas$^e$\\
\hline\\[-0.3cm]
 $M$ $(10^9\,$M$_\odot)^a$& 2 &6\\
 $L_{\rm d}$ $($pc$)^b$  & 150 & 1000$^f$\\
 $\sigma$ $($km$\,$s$^{-1})^c$ & 5 & 12\\
 $E_{\rm kin}$ ($10^{55}\,$erg$\,$s$^{-1})^d$ & 0.5 & 8.6\\
 \hline
\end{tabular}
\end{center}
}

{\footnotesize
$^a$~Total mass of the component. Values from \citet{Ferriere:2001p2879} and \citet{Kalberla:2003p2981}.\\
$^b$~We take twice the observed disk scale height as the true thickness to be considered. \\
$^c$~The parameter $\sigma$ is the 3-dimensional velocity dispersion. We take the 1-dimensional velocity dispersion of the molecular gas to be $2.9\,$km$\,$s$^{-1}$ and of the HI gas to be $6.9\,$km$\,$s$^{-1}$. \\
$^d$~Total kinetic energy of the component, $E_{\rm kin} = 1/2 \, M \sigma^2$.\\
$^e$~The atomic component in principle can be separated into a cold ($T \approx \mbox{few} \times 10^2\,$K) and a hot ($T \approx \mbox{few} \times 10^2\,$K) component. Because they have similar overall distribution we consider them together and take -- whenever possible -- mean values. \\
$^f$~The scale height of HI ranges from $\sim 230\,$pc within $4\,$kpc up to values of $\sim 3\,$kpc at the outer Galactic boundaries. The HI disk therefore is strongly flared. We adopt some reasonable mean value, but note that this introduces additional uncertainty.
}
\label{tab:MW}
\end{table}%

If we use the numbers from Table \ref{tab:MW}, assume $\dot{M}_{\rm in} = \dot{M}_{\rm SF} \approx 3\,$M$_\odot\,$yr$^{-1}$, and adopt the fiducial value $v_{\rm in} = 220\,$km$\,$s$^{-1}$, then Eq.'s~(\ref{eqn:input-MW}) and (\ref{eqn:dissip-MW}) yield
\begin{eqnarray}
\dot{E}_{\rm in} &\approx& + 4.6 \times 10^{40}\,\mbox{erg}\,\mbox{s}^{-1}\,,\\
\dot{E}_{\rm decay} &\approx& - 3.9 \times 10^{39}\,\mbox{erg}\,\mbox{s}^{-1}\,,
\end{eqnarray}
requiring an efficiency of only 
\begin{eqnarray}
\epsilon = |\dot{E}_{\rm decay}| / \dot{E}_{\rm in} \approx 0.08\,.
\end{eqnarray}
In the light of Eq.~(\ref{eqn:density}), this is a reasonable number. If we follow \cite{Heitsch:2009p8345} and adopt densities in the range $n = 0.01$ to $0.1\,$cm$^{-3}$ for the accreting gas clouds and assume a mean ISM density of $~1\,$cm$^{-3}$ in the solar neighborhood \citep{Ferriere:2001p2879} as well as a drop to $~0.1\,$cm$^{-3}$ out at a distance of $25\,$kpc, we expect the efficiency { to be} of order of 10\%. 

\subsection{Spiral Galaxies}
\label{subsec:spirals}

The HI Nearby Galaxy Survey, THINGS, \citep{Walter:2008p2669} opens up the possibility to perform the above analysis for the extended HI disks of other spiral galaxies as well. We obtained the dataset discussed by \citet{Tamburro:2009p3039}, which allows us to analyze HI column density $\Sigma_{\rm HI}$ and vertical velocity dispersion $\sigma_{\rm 1D}$ as function of radius for 11 nearby galaxies. Our sample includes 8 Milky Way type spirals as well as 3 gas-rich dwarfs. Each galaxy map contains between 140.000 and 720.000 data points. The main parameters are provided in Table \ref{tab:THINGS}. Please consult \citet{Tamburro:2009p3039} for further details on the original data set.

\begin{figure*}[tbp]
\vspace*{0.5cm}
\begin{center}
\psfig{file=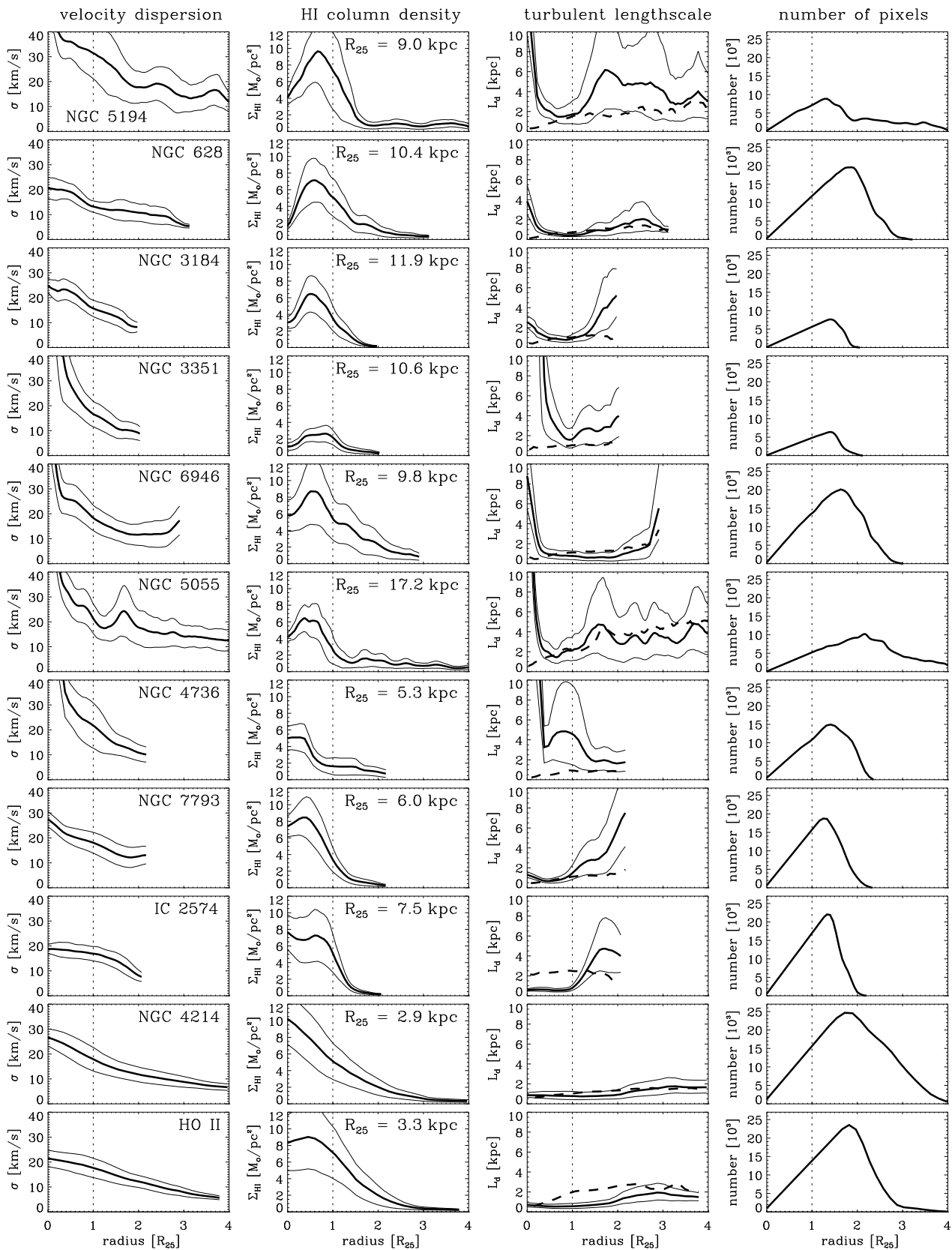,height=21.0cm}
\end{center}
\vspace*{0.2cm}
\caption{Radial distribution of 3D velocity dispersion $\sigma$ (in units of km$\,$s$^{-1}$), HI column density $\Sigma_{\rm HI}$ ($M_\odot\,$yr$^{-1}$), corresponding turbulent scale $L_{\rm d} $ (kpc) taken as twice the disk scale height, computed from hydrostatic balance using Eq.~(\ref{eqn:hd-balance}) (solid lines) and from the total potential using Eq.~(\ref{eqn:total-pot}) (dashed line), number of pixels as function of radius in annuli  of width $\Delta R = 0.5\,$kpc in the 11 THINGS maps. The last column provides an estimate of the statistical significance of the data at each radius. 
The radius is scaled to the optical radius $R_{25}$. The thin lines indicate the 95\% variation. 
}
\label{fig:THINGS}
\end{figure*}

 \begin{figure*}[htbp]
 \vspace*{0.5cm}
\begin{center}
\psfig{file=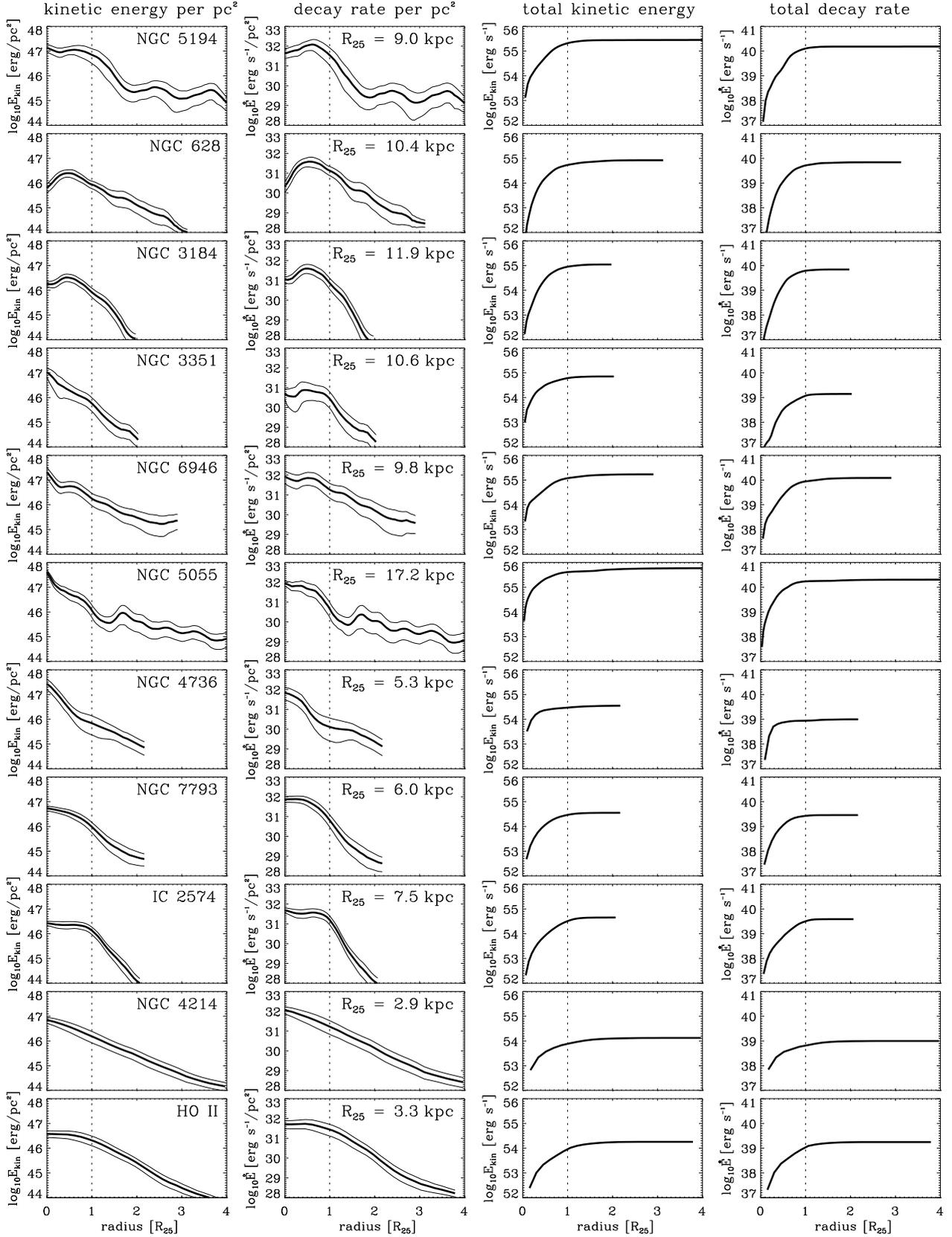,height=21.0cm}
\end{center}
\vspace*{0.2cm}
\caption{Local kinetic energy per unit surface area $E_{\rm kin}$ (erg$\,$pc$^{-2}$), and local turbulent energy decay rate per unit surface area $\dot{E}_{\rm decay}$ (erg$\,$s$^{-1}\,$pc$^{-2}$) using $L_{\rm d}$ derived from vertical hydrostatic balance, Eq.~(\ref{eqn:hd-balance}), together with the cumulative total kinetic energy and cumulative turbulent dissipation rate as function of normalized radius for 11 THINGS galaxies. The thin lines in columns 1 and 2 indicate the 95\% variation at each radius using the scale height derived from Eq.~(\ref{eqn:hd-balance}). 
%
}
\label{fig:energies}
\end{figure*}

\begin{table*}[htdp]
\caption{Observed properties of the analyzed THINGS galaxies.}
\begin{center}
\begin{tabular}{rrccccccccc}
\hline\\[-0.3cm]
& Hubble & $D$ & $v_{\rm rot}$       & $R_{25}$ & $M_{\rm HI}$               & $\dot{M}_{\rm SF}$      & $\langle \sigma \rangle$ & $\langle \Sigma_{\rm HI} \rangle$ &   $\langle L_{\rm d} \rangle_{\rm HD}$ & $\langle L_{\rm d} \rangle_{\rm pot}$ \\
          & type  & Mpc&  km$\,$s$^{-1}$ & kpc     & $10^9\,$M$_{\odot}$ & M$_{\odot}$yr$^{-1}$ & km$\,$s$^{-1}$ & M$_{\odot}$pc$^{-2}$    &kpc       & kpc \\
Name &{\em (1)} & {\em (2)} & {\em (3)} & {\em (4)} & {\em (5)} & {\em (6)} & {\em (7)} & {\em (8)} & {\em (9)} & {\em (10)}  ~\\
\hline  
NGC 5194 &   Sbc  &  8.0 &    220 &   9.0 &    2.5 &   6.05 &  31.69 &   7.18 &   4.72 &   1.32   \\ 
 NGC 628 &   ~Sc  &  7.3 &    220 &  10.4 &    3.8 &   1.21 &  14.28 &   4.55 &   1.00 &    0.80 \\ 
NGC 3184 &   ~Sc  & 11.1 &    210 &  11.9 &    3.1 &   1.43 &  18.34 &   4.48 &   1.46 &    0.99 \\ 
NGC 3351 &   ~Sb  & 10.1 &    200 &  10.6 &    1.2 &   0.71 &  20.89 &   2.05 &   5.42 &     1.01 \\ 
NGC 6946 &   ~Sc  &  5.9 &    200 &   ~9.8 &    4.2 &   4.76 &  18.72 &   5.57 &   1.28 &    1.13 \\ 
NGC 5055 &   Sbc  & 10.1 &    200 &  17.2 &    9.1 &   2.42 &  23.58 &   3.15 &   4.15 &    2.81\\ 
NGC 4736 &   Sab  &  4.7 &    160 &   ~5.3 &    0.4 &   0.43 &  24.98 &   2.99 &   5.11 &     0.71 \\ 
NGC 7793 &   Scd  &  3.9 &    130 &   ~6.0 &    0.9 &   0.51 &  19.64 &   6.06 &   1.40 &     0.91 \\ 
 IC 2574 &   ~Sm  &  4.0 &     80 &   ~7.5 &    1.5 &   0.12 &  17.30 &   5.82 &   1.13 &     2.42 \\ 
NGC 4214 &   Irr  &  2.9 &     60 &   ~2.9 &    0.4 &   0.05 &  16.91 &   4.78 &   1.18 &     1.17\\ 
   HO II &   Irr  &  3.4 &     40 &   ~3.3 &    0.6 &   0.07 &  16.89 &   6.37 &   0.96 &     1.99 \\ 

\hline
\end{tabular}
\end{center}
{\small
{\em (1)} Hubble type as listed in the LEDA data base (URL: http://leda.univ-lyon1.fr/).~ {\em (2)} Distance according to \citet{Walter:2008p2669}.~ {\em (3)} Peak of the rotation curve as obtained from the  Appendix in \citet{deBlok:2008p4057}. Note that our values agree well with the fit formula provided in Appendix B.1 in \citet{Leroy:2008p4217} for galaxies where the flat part of the rotation curve is observed. The only exception is IC~2574, which shows a continuously rising roation curve within the observed radius range.~ {\em (4)} $R_{25}$ is the B-band isophotal radius at 25 mag arcsec$^{-2}$, which is a standard proxy for the optical radius of the galaxy.~ {\em (5)} Total HI mass of the galaxy, obtained by integrating over all pixels, see also Table 4 of \citet{Leroy:2008p4217}.~ {\em (6)} Total star formation rate of the galaxy, as provided by Table 1 of \citet{Walter:2008p2669} or Table 4 of \citet{Leroy:2008p4217}.~ {\em (7)} Surface-density weighted 3-dimensional mean velocity dispersion in the galaxy.~ {\em (8)} Mean HI surface density.~ {\em (9)}  Maximum length scale of the turbulent velocity field. $\langle L_{\rm d} \rangle_{\rm HD}$ is calculated from the disk thickness in each pixel as  $L_{\rm d} = 2 H$ with $H = \sigma_{\rm 1D}^2 / (2 \pi G \Sigma_{\rm HI})$ and then averaged over the entire galaxy using surface-density weighting. Note, that $L_{\rm d}$ varies strongly with radius (see Figure \ref{fig:THINGS}), so that the physical interpretation of average turbulent length scale is not straight forward.~ {\em (10)} Estimate of the mean turbulent dissipation scale $\langle L_{\rm d} \rangle_{\rm pot}$ based on the disk thickness derived from the potential method, Eq.~(\ref{eqn:total-pot}). Note that both approximations give roughly the same numbers with $\langle L_{\rm d} \rangle_{\rm HD} \simgreat \langle L_{\rm d} \rangle_{\rm pot}$ for galaxies with $v_{\rm rot} \simgreat 200\,$km$\,$s$^{-1}$ and $\langle L_{\rm d} \rangle_{\rm HD} \simless \langle L_{\rm d} \rangle_{\rm pot}$ for the dwarfs  with $v_{\rm rot} < 100\,$km$\,$s$^{-1}$.}
\label{tab:THINGS}
\end{table*}%

\begin{table*}[htdp]
\caption{Derived energy input and decay rates and corresponding minimum efficiencies.}
\begin{center}
\begin{tabular}{r@{~~}c@{~~~}c@{~~}c@{~~}c@{~~}c@{~~}c@{~~}c@{~~~~~}c@{~~~}c@{~~~}c@{~~}c}
\hline\\[-0.3cm]
& $\dot{E}_{\rm in}$ & $\dot{E}_{\rm decay,HD}$ & $\epsilon_{\rm HD}$ & $\dot{E}_{\rm decay,pot}$ & $\epsilon_{\rm pot}$ & $\dot{E}_{\rm decay,R_{25}}$ & $\epsilon_{\rm R_{25}}$ & $E_{\rm kin}$  & $f_{>R_{25}}^{E_{\rm kin}}$ & $f_{>R_{25}}^{\dot{E}_{\rm decay,HD}}$  & $\epsilon_{>R_{25}}$\\
& $10^{39}\,$erg$\,$s$^{-1}$ &   $10^{39}\,$erg$\,$s$^{-1}$   & &    $10^{39}\,$erg$\,$s$^{-1}$   & &    $10^{39}\,$erg$\,$s$^{-1}$   &  &  $10^{54}\,$erg & & &\\
Name &{\em (1)} & {\em (2)} & {\em (3)} & {\em (4)} & {\em (5)} & {\em (6)} & {\em (7)} & {\em (8)} & {\em (9)} & {\em (10)} & {\em (11)}~  \\
\hline  
NGC 5194 &     92.39 &      3.41 &    0.037 &       7.08 &    0.077 &       0.45 &   0.005 &    28.20 &      0.26 &      0.17 &    0.006 \\ 
 NGC 628 &     18.48 &      1.97 &    0.106 &       2.06 &    0.111 &       0.06 &   0.003 &     8.39 &      0.38 &      0.28 &    0.027 \\ 
NGC 3184 &     19.90 &      0.82 &    0.041 &       0.96 &    0.048 &       0.04 &   0.002 &    11.03 &      0.21 &      0.13 &    0.005 \\ 
NGC 3351 &      ~8.96 &      0.16 &    0.018 &       1.08 &    0.121 &       0.06 &   0.006 &     ~7.12 &      0.15 &      0.17 &    0.003 \\ 
NGC 6946 &     60.07 &      4.44 &    0.074 &       5.63 &    0.094 &       0.28 &   0.005 &    17.01 &      0.32 &      0.29 &    0.020 \\ 
NGC 5055 &     30.54 &      1.56 &    0.051 &       3.59 &    0.117 &       0.15 &   0.005 &    58.98 &      0.31 &      0.15 &    0.008 \\ 
NGC 4736 &      ~3.47 &      0.52 &    0.151 &       6.42 &    1.848 &       0.29 &   0.083 &     ~3.52 &      0.17 &      0.12 &    0.018 \\ 
NGC 7793 &      ~2.72 &      1.87 &    0.686 &       1.95 &    0.716 &       0.13 &   0.050 &     ~3.57 &      0.17 &      0.07 &    0.047 \\ 
 IC 2574 &      ~0.24 &      2.18 &    9.009 &       0.60 &    2.459 &       0.10 &   0.417 &     ~4.52 &      0.28 &      0.18 &    1.600 \\ 
NGC 4214 &      ~0.06 &      1.28 &   $\!\!$22.457 &       1.17 &   $\!\!$20.549 &       0.21 &   3.653 &     ~1.33 &      0.49 &      0.42 &    7.831 \\ 
   HO II &      ~0.04 &      1.77 &   $\!\!$50.070 &       0.67 &   $\!\!$19.065 &       0.17 &  4.900 &     ~1.81 &      0.59 &      0.50 &   $\!\!\!$18.554 \\ 
\hline
\end{tabular}
\end{center}
{\small
{\em (1)}  Total kinetic energy provided by infalling gas, calculated from Eq.~(\ref{eqn:dissip1}) using $\dot{M} = \dot{M}_{\rm SF}$ and $v_{\rm in} = v_{\rm rot}$.~ {\em (2)} Total dissipation rate of turbulent kinetic energy $\dot{E}_{\rm decay,HD}$ obtained from integrating Eq.~(\ref{eqn:dissip}) over the entire galaxy using $L_{\rm d}$ derived from vertical hydrostatic equilibrium.~ {\em (3)} Minimum efficiency $\epsilon_{\rm HD} = |\dot{E}_{\rm decay,HD}|/\dot{E}_{\rm in}$ required for the conversion of infall motion into turbulent energy in the disk.~ {\em (4)}  Integrated turbulent decay rate $\dot{E}_{\rm decay,pot}$ based on the  potential method.~ {\em (5)} Corresponding minimum efficiency $\epsilon_{\rm pot}$.~  {\em (6)}  Integrated turbulent decay rate $\dot{E}_{\rm decay,R_{25}}$ based on the assumption that the outer scale of the turbulent velocity field  is equal to the size of the disk $L_{\rm d} \sim2\,R_{25}$. We list this value to provide an estimate of the uncertainty introduced by not knowing $L_{\rm d}$ very well. In Section \ref{subsec:uncertainties} we argue that  $\dot{E}_{\rm decay,R_{25}}$ is a strict lower limit and that the true decay rate  probably lies more closely to $\dot{E}_{\rm decay,HD}$.~ {\em (7)} The corresponding required minimum efficiency $\epsilon_{\rm R_{25}}$ for accretion driven turbulence to work.~  {\em (8)}  Total kinetic energy $E_{\rm kin}$ integrated over the whole galaxy.~  {\em (9)} Fraction of total kinetic energy outside the optical radius $R_{25}$. ~  {\em (10)} Fraction of turbulent energy decay rate outside of $R_{25}$ using the disk vertical scale height from hydrostatic balance. The total rate is given in column (2).~  {\em (11)} Corresponding required accretion efficiency to drive the turbulence in the outer disk for $R > R_{25}$. Compare to the total galactic value given in column (3). The outer disk value is typically by a factor of 5 lower.
}
\label{tab:efficiencies}
\end{table*}%

{ 
We integrate over the entire map to get the total HI mass, and convert $\sigma_{\rm 1D}$ to the 3-dimensional velocity dispersion $\sigma$ assuming isotropy. We read off the measured star formation rate, $\dot{M}_{\rm SF}$, from Table 1 of \citet{Walter:2008p2669},  and consider the rotation curves from \citet{deBlok:2008p4057}, where we adopt the peak value $v_{\rm rot}$ for our analysis. An estimate of the turbulent length scale $L_{\rm d}$ is more difficult to obtain. We follow \citet{Leroy:2008p4217} and derive an estimate of the thickness $H$ of the HI layer by assuming hydrostatic equilibrium at every pixel. In this case, 
\begin{equation}
\label{eqn:hd-balance}
H \approx \sigma_{\rm 1D}^2 / (2 \pi G \Sigma_{\rm HI})\,, 
\end{equation}
with $\sigma_{\rm 1D} =  3^{-1/2}\sigma$ and    $G$ being the 1-dimensional velocity dispersion and the gravitational constant, respectively. 
Note, that we neglect the contribution from molecular gas  as well as from stars in the determination of $H$ which could be significant especially in the inner regions of the disk.  In principle we would need to calculate  $H \approx \sigma_{\rm 1D}^2 / (2 \pi G \Sigma_{\rm tot})$ with $\Sigma_{\rm tot} =  \Sigma_{\rm HI} + \Sigma_{\rm H_2} +  \Sigma_{\star}$ being the combined surface density of gas and stars. Our estimate of the scale height $H$ in the inner parts of the galaxy therefore is an upper limit. 
An alternative estimate for the disk thickness is based on the total enclosed mass at any given radius $R$ using the local circular velocity $v_{\rm rot}$ (see Appendix \ref{app:potential-method}),
\begin{equation}
\label{eqn:total-pot}
H \approx R \,\sigma_{\rm 1D}/ v_{\rm rot} \,.
\end{equation}
Again, we follow \citet{Leroy:2008p4217}  and approximate the rotation curve with the fit formula,
\begin{equation}
v_{\rm rot} (R) = v_{\rm flat} \left[ 1 - \exp\left(-\frac{R}{R_{\rm flat}}\right)\right]\,,
\end{equation}
using the values $v_{\rm flat}$ and $R_{\rm flat}$ for the flat parts of the rotation curve from their Table 4. We point out, that except maybe for the inner disk the mass distribution is dominated by the dark matter content of the galaxy. 

Once we have calculated $H$ at each location in the map, we set the local turbulent scale length to $L_{\rm d} =  2\,H$. For most galaxies in the sample, both of the above estimates lie within a factor of two or less of each other, with the potential method usually giving somewhat lower numbers for the large spirals and higher values for the dwarf galaxies. The exceptions are NGC~5193, NGC~3351, NGC~5055, and NGC~4736 which  have a strong molecular component in the center \citep{Leroy:2008p4217} and  where our $L_{\rm d}$ estimate based on vertical hydrostatic balance using $\Sigma_{\rm HI}$ consequently is too large. NGC~5055 and NGC~4736,  furthermore, are characterized by extended HI streamers at $\sim R_{25}$ which results in a locally enhanced velocity dispersion \citep{Walter:2008p2669}, again leading to inflated $L_{\rm d}$ values from hydrostatic balance.     
For comparison with the Milky Way, mean velocity dispersion $\langle \sigma \rangle$, mean HI surface density $\langle \Sigma_{\rm HI} \rangle$, and mean turbulent scale $\langle L_{\rm d} \rangle$ based on both methods are obtained for each galaxy as surface density weighted average over all pixels and are provided in Table \ref{tab:THINGS} as well.

The radial variation of $\sigma$, $\Sigma_{\rm HI}$, and $L_{\rm d}$ is shown in Figure \ref{fig:THINGS}. In addition, this figure also indicates the number of pixels contributing to radial annuli of width $\Delta R = 0.5\,$kpc for each of the THINGS maps. This provides some estimate of the statistical significance of the data at different radii. We notice that outside of $R_{25}$, which is a proxy for the optical radius of the stellar disk, the values of $\sigma$ typically drop below $\sim 20\,$km$\,$s$^{-1}$ and can get as low as $\sim 5\,$km$\,$s$^{-1}$ in the outer disk of  the dwarf galaxies in the sample, with statistical uncertainties of  $2 - 3\,$km$\,$s$^{-1}$.  Inside of $R_{25}$ the velocity dispersion can be significantly higher. Here stellar sources provide additional energy for driving turbulent motions \citep{MacLow:2004p2713}. The HI surface density also drops outside of $R_{25}$. We note, however, that for some galaxies with very extended disks, $\Sigma_{\rm HI}$ can remain as high as $1\,$M$_{\odot}\,$pc$^{-1}$ out to $4\,R_{25}$. We also point out, that some galaxies reveal a noticeable HI depletion in their central regions where the gas is mostly molecular \citep{Leroy:2008p4217, Leroy:2009p4218}. Many galaxies in our sample show significant flaring in the outer disk. Some galaxies also show large $L_{\rm d}$ values in the very inner parts. Recall that this is in part an artifact of our analysis, because we neglect the contribution to the graviational potential from the stellar disk and from the molecular gas when computing the scale height from vertical hydrostatic balance. Inflated central $L_{\rm d}$ therefore correlate well with the inner depletion of $\Sigma_{\rm HI}$.

\begin{figure}[t]
\hspace*{0.3cm}\centerline{\psfig{file=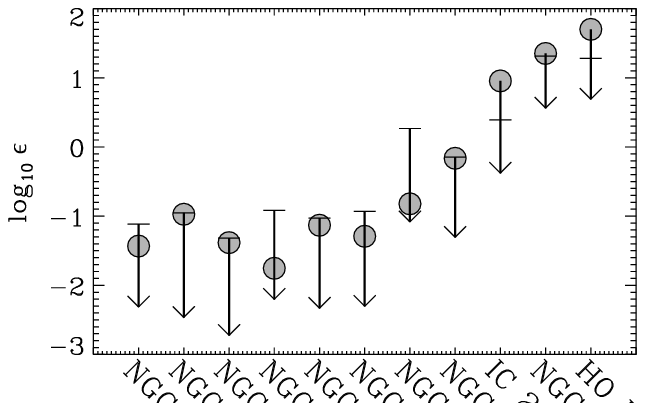,width=8.5cm}}
\vspace*{0.8cm}
\caption{Minimum efficiency  required to sustain the observed level of turbulence by accretion from the galactic halo for the sample of THINGS galaxies. Our fiducial values, $\epsilon_{\rm HD}$ (filled circles) are based on using the disk thickness from vertical hydrostatic balance as outer scale of the turbulent velocity field $L_{\rm d}$. To give an illustration of the uncertainties involved in estimating $L_{\rm d}$, we also show $\epsilon_{\rm pot}$ derived from using the global potential to calculate the disk thickness (horizontal symbol), and $\epsilon_{\rm R_{25}}$ where we take the diameter of the optical disk as proxy for $L_{\rm d}$ (arrow down). The latter is a strict lower limit, with the true minimum efficiencies probably being closer to our fiducial estimate. Note, that for the dwarf galaxies in our sample ($v_{\rm rot}  < 100\,$km$\,$s$^{-1}$) $\epsilon_{\rm HD} \simgreat \epsilon_{\rm pot}$ while the opposite is true for the Milky Way type galaxies   ($v_{\rm rot}  \simgreat 200\,$km$\,$s$^{-1}$).}
\label{fig:epsilon-range}
\end{figure}

With this information, we can now compute the local turbulent kinetic energy as well as the local kinetic energy decay rate based on the two estimates of the disk scale height. We display the result in the left two columns of Figure \ref{fig:energies}. For simplicity, we only show the decay rate based on the assumption of vertical hydrostatic equlibrium. We integrate over all radii to obtain the total decay rate $\dot{E}_{\rm decay}$ and, once again assuming steady state, take the star formation rate $\dot{M}_{\rm SF}$ to estimate $\dot{E}_{\rm in}$ from Eq.~(\ref{eqn:dissip-MW}). This allows us to calculate the minimum efficiency $\epsilon = |\dot{E}_{\rm decay}|/\dot{E}_{\rm in}$ needed for sustaining the observed disk turbulence by gas accretion. The results are presented in Table \ref{tab:efficiencies} and graphically illustrated in Figure \ref{fig:epsilon-range}. 

We notice that all galaxies similar to the Milky Way, i.e.\ those with Hubble types ranging from Sb to Sc with $v_{\rm rot} \simgreat 200\,$km$\,$s$^{-1}$, require efficiencies of only 10\% or less. This holds despite variations in star formation rate or total gas mass of almost a factor of ten. As in Section \ref{subsec:MW} we argue that $\epsilon \simless 0.1$ is a reasonable number. Given that the true turbulence dissipation scale could exceed the disk thickness and potentially be as large as the disk diameter (see Section \ref{subsec:uncertainties} below), the derived decay rates are upper limits and the minimum efficiency of accretion driven turbulence might be considerably smaller. We point out that the model clearly fails for the dwarf galaxies in our sample with $v_{\rm rot} < 100\,$km$\,$s$^{-1}$. The kinetic energy added by accretion is not sufficient to compensate for the energy loss by turbulent decay. Either these galaxies accrete more mass than inferred from their low star-formation rates, or there are other processes that dominate the disk turbulence on all scales. Both explanations appear equally likely and ask for more detailed studies with specific focus on dwarf galaxies.

\begin{figure*}[t]
\hspace*{0.3cm}\centerline{\psfig{file=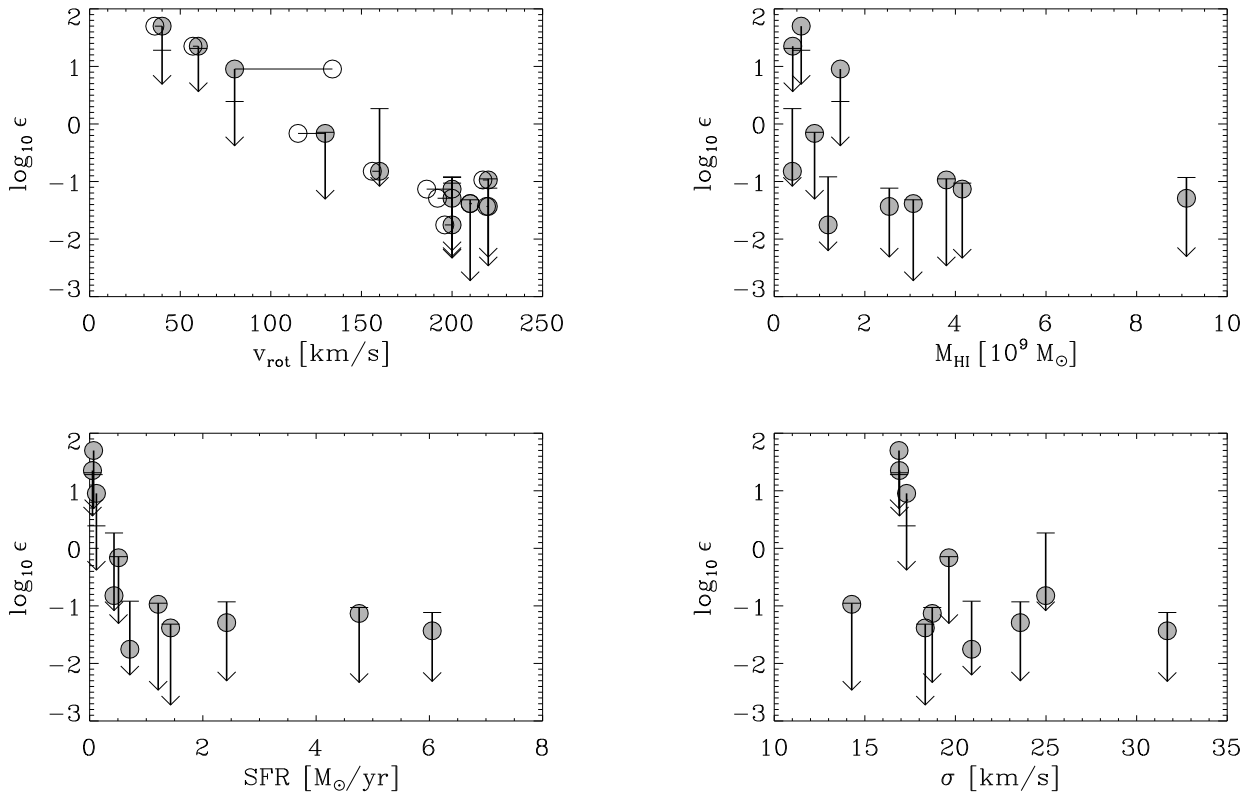,width=18.0cm}}
\vspace*{0.2cm}
\caption{Correlation between required minimum efficiency $\epsilon$ and rotational velocity $v_{\rm rot}$ {\em (top left)}, total HI mass $M_{\rm HI}$ {\em (top right)}, star formation rate {(bottom left)}, and average 3-dimensional velocity dispersion $\sigma$ {\em (top right)}. The used symbols are the same as in Figure  \ref{fig:epsilon-range}. Note that in the top left plot, we use open circles to denote the shift introduced when using the fit parameter $v_{\rm flat}$ \citet{Leroy:2008p4217}  instead of the observed maximum of the rotation curve (Table {tab:THINGS}). This is only significant for the galaxy  IC~2574 which has a continuously rising roation curve within the observed radius range. The efficiency $\epsilon$ correlates well only with $v_{\rm rot}$.}
\label{fig:correlations}
\end{figure*}

In order to better understand the physical origin of this variations, we plot the minimum efficiency values obtained above as function of different galaxy parameters in Figure \ref{fig:correlations}. We see that $\epsilon$ correlates very well with the rotational velocity $v_{\rm rot}$ of the galaxy, $\log_{10} \epsilon \approx 2.3 - v_{\rm rot}/(60\,$km$\,$s$^{-1})$. This is understandable because we use $v_{\rm rot}$ as proxy for the infall velocity and hence for the kinetic energy of the infalling material. The rotational velocity is a good measure of the total mass including dark matter and stellar component. If we exclude the dwarf galaxies in the sample, we find no correlation between $\epsilon$ and total HI mass, star formation rate, and velocity dispersion. This is somewhat surprising. We had expected the efficiency to scale inversely with HI mass, because the total amount of turbulent energy that needs to be replenished scales linearly with $M_{\rm HI}$ assuming that $\sigma$ is roughly constant. The contrary is the case, if we focus on the  dwarf galaxies we see that they have high $\epsilon$ values despite small $M_{\rm HI}$. Because the decay rate, Eq.~(\ref{eqn:dissip}), scales with the third power of the velocity dispersion, we had also expected that $\epsilon$ strongly depends on $\sigma$. Again, this is not seen. From our small sample of 11 galaxies, we conclude that the total mass of the galaxy is the main parameter determining the potential importance of accretion driven turbulence.  It would be interesting to perform a similar analysis with a larger sample of galaxies to have better statistics.
}

{ 
\subsection{Main Uncertainties}
\label{subsec:uncertainties}

The processes discussed here are subject to large uncertainties. First of all, virtually all quantities that enter the theory vary with radius. Defining a galaxy-wide average value is not a trivial task. In cases where we have extended HI maps, this is not a problem, because we can integrate over all radii and obtain well defined global values for $E_{\rm kin}$, $\dot{E}_{\rm decay}$, and $\dot{E}_{\rm in}$. However, for galaxies with only a few pixels across or for observations with insufficient sensitivity to detect the extended disk the errors can be considerable. Second, we do not know how the gas enters the galaxy. Does it fall onto the outer disk in discrete cold streams \cite[as indicated by cosmological simulations at high redshift][]{Dekel:2009p1173,Agertz:2009p1291}? Or does it condense out of the tenuous halo gas and enter the disk more gently and more distributed as proposed by \citet{Peek:2009p9660}? Both processes could lead to  very different efficiencies. 

The third and probably most severe uncertainty concerns the turbulent length scale. The observational data are not very conclusive, with estimates of the turbulent scale ranging from only $4\,$pc \citep{Minter:1996p16767} from scintillation measurements in the Milky Way up to $6\,$kpc \citep{Dib:2005p16777} from computing the autocorrelation length of HI in Holmberg II with large uncertainties in both values. Very careful statistical analyses of the power spectrum in several nearby molecular clouds  by  \citet{Ossenkopf:2002p334} as well as  \citet{Brunt:2003p17014} and \citet{Brunt:2009p16966}, however, indicate that the bulk of the turbulent energy is always carried by the largest scales observed. This is consistent with turbulence being driven from the outside. 

Throughout most of this paper we assume that the outer scale of the turbulent cascade is comparable to the disk thickness, i.e. we take it as being twice the vertical scale height. We argue that this is the relevant upper length scale in the system. Only then can we speak of 3-dimensional turbulence, where we have good estimates of the decay properties  \cite[e.g.][]{MacLow:1998p384,Stone:1998p12501,Padoan:1999p516,MacLow:1999p251,Elmegreen:2000p9513}. One may propose, however, that the turbulent cascade extends all the way across the galactic disk. In this case, $L_{\rm d} \sim 2\,R_{25}$. Turbulence is mostly 2-dimensional and it is not clear how to estimate its decay properties in the differentially rotation disk. 
{  It has been speculated, however, that the decay properties of 2- and 3-dimensional turbulent flows could be equivalent as long as they are in the strongly supersonic regime, because dissipation occurs mostly in sheet-like shocks \citep{AvilaReese:2001p13326}.
}
If Eq.~(\ref{eqn:dissip}) is still approximately correct, the decay rate is considerably lower because $R_{25} \gg H$, and our model can tolerate even lower minimum efficiencies (Table \ref{tab:efficiencies}).  On the other hand, the velocity difference across the disk is of order of the rotational velocity $v_{\rm rot}$, which for the Milky Way type galaxies greatly exceeds $\sigma$. Because $H/R \sim \sigma/v_{\rm rot}$, Eq.~(\ref{eqn:total-pot}), it is likely that both effects approximately cancel and that the decay rates based on $H$ and on $R_{25}$ are comparable. 
{  This is in fact what we would expect from a self-similar turbulent cascade, where the behavior is determined by the physical properties at the dissipation scale.
}
We adopt $L_{\rm d}$ based on the disk scale height obtained from vertical hydrostatic balance, Eq.~(\ref{eqn:hd-balance}), as our fiducial value. The range of $\epsilon$ values for each galaxy in our sample associated with the uncertainties in $L_{\rm d}$  is  illustrated in Figure \ref{fig:epsilon-range}. 

We also want to call attention again to the fact that we have neglected the molecular gas content in our analysis of the THINGS galaxies. Indeed, some of these galaxies contain an appreciable amount of molecular gas \citep{Leroy:2009p4218}. However, this gas is mostly contained within $R_{25}$ and in addition carries only little turbulent kinetic energy compared to the atomic component (see also Section \ref{subsec:MW} for the Milky Way). The error involved in focusing on atomic gas only is therefore small.
}

\subsection{Outer Disks of Spiral Galaxies}
\label{subsec:outer-disks}

It is well know now that the gaseous disk extends well beyond the optical radius \citep{Thilker:2007p5194,Zaritsky:2007p10073}. The sample of THINGS galaxies allows us to study the turbulent energy content of extended disks in more detail. { We begin by asking what parts of the disk carry most of the turbulent kinetic energy. We calculate the cumulative kinetic energy as function of radius as well as the cumulative energy decay rate and plot both quantities columns 3 and 4 of Figure \ref{fig:energies}. We see that a significant fraction (between 15\% and 59\%) of the total kinetic energy is carried by the outer disk. The numbers together with the required accretion efficiency are provided in Table \ref{tab:efficiencies}. This finding has important consequences for our understanding of the origin of this turbulence. While within $R_{25}$ energy and momentum input from stellar sources  (supernovae, stellar winds and outflows, expanding HII regions) can contribute significantly to driving ISM turbulence, this approach fails for the outer parts.} Here accretion from the extended gaseous halo, maybe in concert with the magneto-rotational instability \citep{Tamburro:2009p3039}, is the only astrophysical driving source available. The required efficiencies for Milky Way type galaxies are about 1\%. This is a very low value and we conclude that accretion could easily drive the turbulence in the outer disk  of present-day spiral galaxies. 


\subsection{Clumpy Galaxies at High Redshifts}
\label{subsec:high-z}
The mechanism that we propose here to drive internal turbulence is very generic and likely to operate on many different spatial and temporal scales. High redshift galaxies, for example such as detected in the Hubble Ultra Deep field, are observed to be very irregular, with blue clumpy structure, asymmetry, and a lack of central concentration \citep{Elmegreen:2005p4432,Conselice:2003p4785,Elmegreen:2009p4347}. They are typically characterized by a considerably higher degree of internal turbulence, e.g.\ as reflected by the large observed line width of H$\alpha$ emission, than present-day galaxies in the same mass range \citep{Genzel:2008p4656}. There seems to be an evolutionary trend with decreasing redshift from clumpy galaxies with no evidence of interclump emission to those with faint red disks. This trend continues to present-day spiral galaxies of ßocculent or grand design types  \citep{Elmegreen:2009p4347}. Some clumpy galaxies at high redshift resemble massive versions of local dwarf irregular galaxies. They exhibit very high gas fractions and appear to be in a very young evolutionary state. Soon after their discovery it has been recognized that their strong turbulence is difficult to maintain { with} internal sources as the stellar feedback processes that act in present-day galaxies are relatively ineffective when the velocity dispersion of the whole interstellar medium is large \citep{Elmegreen:2009p4400}. Instead it has been proposed \cite[e.g.][]{Genzel:2008p4656,Elmegreen:2009p4347} that this high degree of turbulence could be driven by cold accretion streams as found in numerical simulations with detailed treatment of the thermodynamic behavior of the infalling gas \citep{Birnboim:2003p4746, Semelin:2005p4771,Dekel:2006p4748,Agertz:2009p1291,Ceverino:2009p1271}.

If we assume the accretion onto the clump at any instance in time is driven by the clump's self-gravity, then $1/2 v_{\rm in}^2 \approx GM/L_{\rm d}$ and Eq.~(\ref{eqn:mdot}) simplifies to 
\begin{eqnarray}
\dot{M}_{\rm in} \approx \frac{1}{\epsilon} \frac{\sigma^3}{2 G}\,.
\end{eqnarray}
With typical clump sizes $L_{\rm d} \approx 1\,$kpc, masses of about $\sim 10^8\,$M$_{\odot}$, and 3-dimensional velocity dispersions $\sigma \approx 30\,$km$\,$s$^{-1}$, the required accretion rate to drive the clump's internal turbulence is 
\begin{eqnarray}
\dot{M}_{\rm in} \approx \frac{1}{\epsilon}\,1\,\mbox{M}_\odot\,\mbox{yr}^{-1}\,.
\end{eqnarray}
The observed star formation rates of clumpy galaxies lie in the range $10 - 50\,$M$_{\odot}$yr$^{-1}$. If we assume that these numbers can be used as proxy for the mass accretion rate then we see that again efficiencies of order of 10\% or less are sufficient to drive the clumps internal turbulence. This is consistent with the $\epsilon$-values estimated from numerical experiments as discussed in Section \ref{subsec:input-rate}, and adds further support for the hypothesis of accretion driven turbulence in clumpy galaxies \cite[see also][]{Elmegreen:2009p15769}.

{ 
\subsection{Further Discussion}
\label{subsec:limits}

The analysis above relies on the assumption that galaxies evolve in steady state so that the star-formation rate is matched by the infall of fresh gas from an extended halo. For most present-day field galaxies this appears to be a reasonable assumption. However, it breaks down for highly perturbed systems, e.g.\ when galaxies experience a major merger or are tidally disturbed in the central regions of dense galaxy clusters. Strong perturbations lead to enhanced star formation without being necessarily accompanied by additional gas infall. It is the original disk gas that is converted into new stars at increased rate. We therefore expect our model to work best for disk galaxies that are marginally unstable and where stellar birth proceeds in a self-regulated fashion similar to the Milky Way.  Indeed the $L_\star$ galaxies in our sample with $v_{\rm rot} \approx 200\,$km$\,$s$^{-1}$ all fit our model very well, while the dwarf galaxies with $v_{\rm rot} \simless 100\,$km$\,$s$^{-1}$ do not. These galaxies are characterized by very irregular clumpy structure. They resemble the clumpy galaxies seen at high redshift \citep{Elmegreen:2005p4432,Conselice:2003p4785}. Taken at face value, our simple equilibrium model therefore should not apply to those galaxies either, as they most likely are still in the phase of rapid mass growth through merging and accretion of massive cold clouds \citep{Genzel:2008p4656,Elmegreen:2009p4347}. However, we can apply our model to structures {\em within} these galaxies and speculate that the turbulence observed within the dense clumps is driven by accretion (Section \ref{subsec:high-z}). This is similar to the mechanism we propose in Section \ref{sec:clouds}  to drive turbulence in present-day molecular clouds in the Milky Way.

We also need to point out, that our analysis by no means implies that other sources of turbulence are not important. Clearly in the inner parts of the disks of $L_\star$-type galaxies, stellar feedback plays a key role and can provide sufficient energy to drive the observed turbulence \citep{MacLow:2004p2713}. In addition, it is difficult to see how the accretion energy, that we expect due to angular momentum conservation to be mostly added to the outer regions of the disk, is able to reach the inner disk within reasonable timescales and without being dissipated. If associated with net mass transport, one would expect a mean inward flow of order of $5\,$km$\,$s$^{-1}$, which is not observed  \cite[Wong et al.\ 2004, however, see ][ for an alternative accretion scenario]{Peek:2009p9660}. It is important to note further, that the accretion of halo gas is likely to be non-spherical and may excite wave-like perturbations that then could tap to the system's rotational energy. This provides an additional reservoir for driving turbulence in the disk so that even small accretion rates could lead to effective disk heating. 

Furthermore, we note that our analysis neglects the influence of galactic fountain flows. Expanding supernova bubbles or large HII regions could transport hot and metal enriched material into the halo, where it cools and eventually falls back onto the galactic disk \cite[][]{Corbelli:1988p18791}. This process is a key element of the matter cycle and enrichment history on global galactic scales \cite[][]{Spitoni:2009p18779}. In addition, the kinetic energy associated with the returning material could contribute to driving ISM turbulence just like infalling fresh material does. This supplementary source again renders the above estimate of the minimum efficiency required for accretion-driven turbulence to work an upper limit. In particular, galactic fountains should not affect the outer disk much, because they deliver the ejected material close to the radius where it originates from. They are important only for the inner, star-forming parts of the disk \cite[][]{Melioli:2008p18789,Melioli:2009p18793}.
}

\section{Molecular Cloud Turbulence}
\label{sec:clouds}

\subsection{Theoretical Considerations}
\label{subsec:MC-theory}

Since the seminal work of \citet{Larson:1981p12875}, it is well established that molecular clouds are highly turbulent. 
The 3-dimensional velocity dispersion in these objects varies with their size, $L$, and typically follows the relation,
\begin{eqnarray}
\sigma \approx 0.8 \; {\rm km \, s}^{-1} \left( { L \over 1 {\rm pc} } \right)^{0.5}\,.
\label{sigma}
\end{eqnarray}
The physical origin of this turbulence is not fully understood yet. { In particular, the question as to whether it is 
injected from the outside, e.g.\ by colliding flows \cite[][]{Hunter:1986p18192,Vishniac:1994p18335},}
 or driven by internal sources such as protostellar outflows \citep{Li:2006p13226,Banerjee:2007p6178,Nakamura:2008p13041,Wang:2010p23176} or expanding HII regions  \citep{Matzner:2002p13240} or supernovae \citep{MacLow:2004p2713}, is still subject to considerable debate. 
We favor the first assumption as observations indicate that molecular cloud turbulence is always dominated by the largest-scale
modes accessible to the telescope \citep{Ossenkopf:2002p334,Brunt:2003p17014,Brunt:2009p16966}. In addition, the amount of turbulence in molecular clouds with no, 
or extremely low star formation like the Maddalena cloud or the Pipe nebula, is significant and broadly comparable 
to the level of turbulence observed towards star forming clouds. Both facts seem difficult to reconcile with turbulence being 
driven from internal stellar sources. 

We argue that it is the very process of cloud formation that drives its internal 
motions by setting up a turbulent cascade that transports kinetic energy from large to small scales in a universal and 
self-similar fashion \citep{Kolmogorov:1941p10448}. Our hypothesis is that molecular clouds form at the 
{ stagnation points of large-scale convergent 
flows \cite[e.g.][]{BallesterosParedes:1999p13248,Hartmann:2001p17469,Klessen:2005p15,Heitsch:2006p12598,Hennebelle:2008p832,Banerjee:2009p6188}, maybe triggered by spiral density waves or other global perturbations of the gravitational potential.} As the density goes 
up the gas can cool efficiently, turn from being mostly atomic to molecular, and shield itself from the external radiation field. 
As long as the convergent flow continues to deliver fresh material the cloud grows in mass and is confined  by the combined thermal and ram pressure of the infalling gas. Because the molecular gas is cold its internal turbulent motions are strongly supersonic. Consequently, the cloud develops a highly complex morphological and kinematic structure with large density contrasts. Some of the high-density regions become gravitational unstable and go into collapse to form stars. This modifies the subsequent evolution, as stellar feedback processes now contribute to the energy budget of the cloud. 

\begin{figure}[tbp]
\centerline{\psfig{file=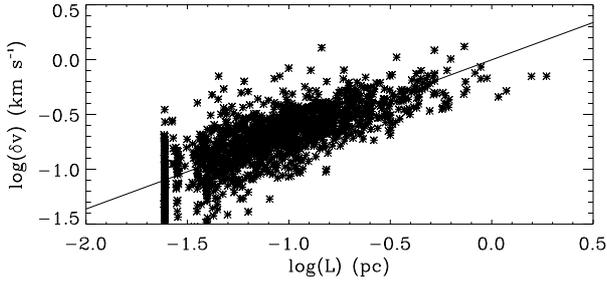,width=8.5cm}}
\caption{Internal velocity dispersion of clumps produced in colliding flows simulations.}
\label{velo_disp}
\end{figure}

\begin{figure}[htbp]
\centerline{\psfig{file=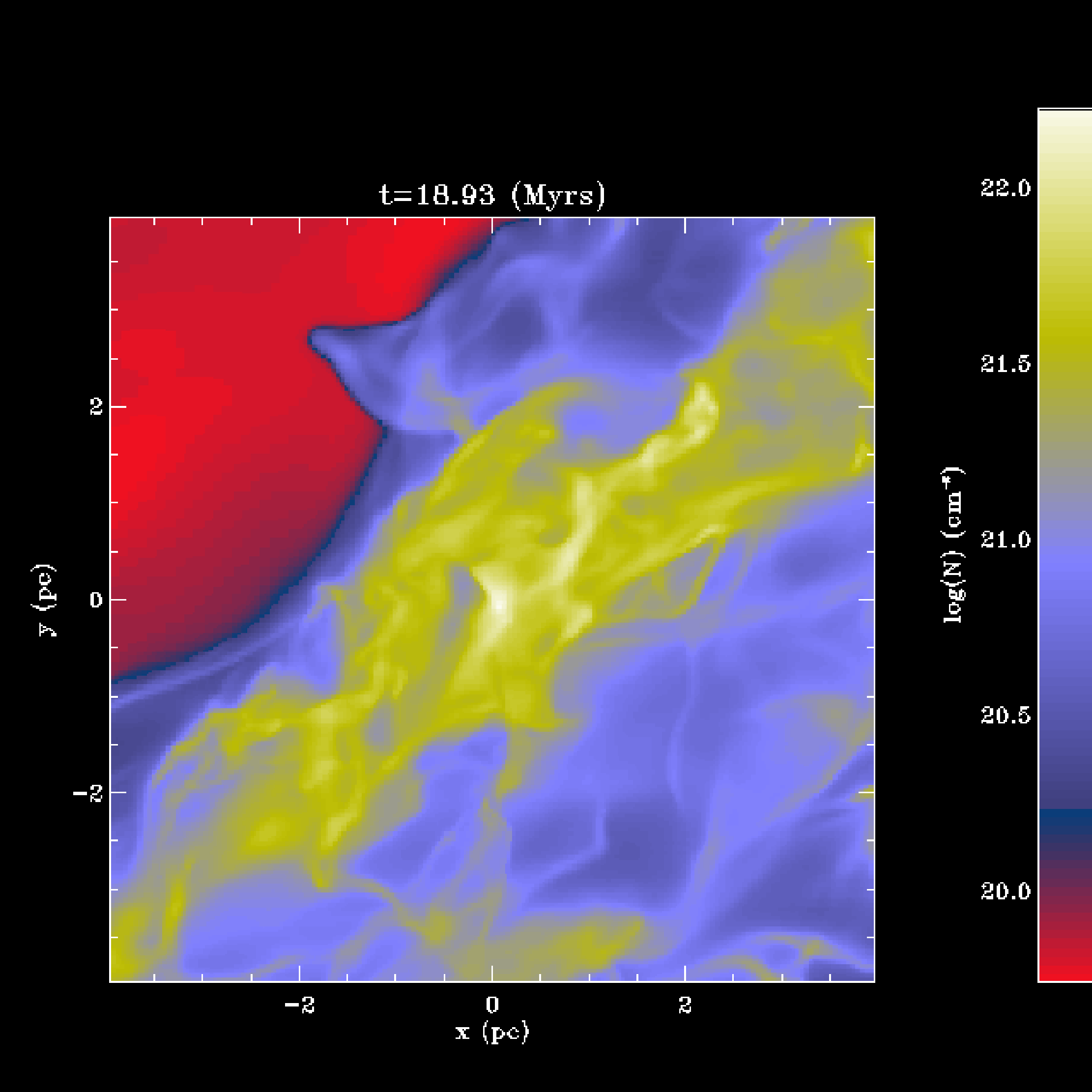,width=8.5cm}}
\caption{Column density for a  clump produced in the simulation.}
\label{column_dens}
\end{figure}

{ A long series of numerical simulations focusing on molecular cloud dynamics and attempting to build up these clouds from diffuse gas at the stagnation points of convergent larger-scale flows have indeed  shown that accretion can sustain a substantial degree of  turbulence in the newly formed cloud \citep{Walder:1998p12610,Walder:2000p17609,Koyama:2002p17620,Folini:2006p12636,Heitsch:2006p8314,VazquezSemadeni:2003p13253,VazquezSemadeni:2006p13319,VazquezSemadeni:2007p47,Hennebelle:2008p832,Banerjee:2009p6188}.} Figure~\ref{velo_disp} shows the internal velocity dispersion of molecular cloud clumps  extracted from the high resolution numerical simulation  described in Section  \ref{subsec:input-rate} as a function of their size. 
These clumps  are defined by a simple clipping algorithms, selecting cells with densities larger than 2500 cm$^{-3}$. The internal velocity  dispersion is then  computed by computing the rms velocity with respect to the cloud  bulk velocity.  As can be seen, the velocity dispersion is compatible with Larson's relation, Eq.~(\ref{sigma}).
To further illustrate this,  Fig.~\ref{column_dens} shows 
the column density of one of the clumps formed in the simulation.
These simulations suggest that continuous accretion of diffuse material in a molecular cloud
is sufficient to maintain a high level of turbulence inside the cloud. 

\subsection{Application to Molecular Clouds in the LMC}
\label{subsec:MC-LMC}

{ 
The best evidence for gas accretion inside molecular clouds may be given by observations in the Large Magellanic Cloud (LMC) reported by \citet{Blitz:2007p13433}, \citet{Fukui:2009p13386}, and \citet{Kawamura:2009p18822}. These authors distinguish 3 types of giant molecular clouds that they interpret as an evolutionary sequence. During the first phase, which should last about 6$\,$Myr based on statistical counting,  the clouds are not forming massive stars and therefore exhibit a low star formation rate. During the second phase, which may last about 13$\,$Myr, massive stars form but not clusters. The last phase is characterized by  the presence of both massive stars and clusters. 
While the mean mass of the clouds observed in the first phase is of the order of $1.1 \times 10^5\, {\rm M}_\odot$ the mean mass of the clouds in the second phase is about $1.7 \times 10^5 \,{\rm M}_\odot$.
This implies that the giant molecular clouds in the LMC are on average accreting at a rate of about 
\begin{eqnarray}
\dot{M} \approx 10^{-2} \,{\rm M}_{\odot} \; {\rm  yr}^{-1}.
\label{mdot}
\end{eqnarray}
Note that \cite{Fukui:2009p13386} quote a slightly larger value of about $5 \times 10^{-2} \;{\rm M}_{\odot} \; {\rm  yr}^{-1}$ and note also that similar values are found in numerical simulations of molecular cloud formation in convergent flows \cite[see, e.g.~Figure 7 in][]{VazquezSemadeni:2009p17155}. In the following discussion, we adopt our estimate as fiducial value, but keep in mind the alternative higher estimate.
}

The  amount of energy which is delivered by this process as well as 
the energy dissipated per unit time in the cloud are given by Eq.'s.~(\ref{eqn:dissip}) and (\ref{eqn:dissip1}). 
{  To obtain an estimate for the relevant parameters, 
we take the outer scale of the turbulence, $L_{\rm d}$, to be equal to the size of the molecular cloud, and adopt the observed relation between cloud mass and size \cite[see, e.g., Figure 1 in][]{Falgarone:2004p13462},
\begin{eqnarray}
\label{eqn:larson2}
M \approx  10 \,{\rm M}_\odot  \left( { L \over 1 {\rm pc}}  \right)^{2.3}.
\end{eqnarray}
Note, that this empirical behavior implies that the cloud has a fractal dimension of $2.3$ or, more or less equivalently, has large internal density contrasts. Recall that the scaling relation for homogeneous clouds is simply $M \propto L^3$.} 
We then calculate the infall velocity $v_{\rm in}$ using mass conservation.
The accretion rate $\dot{M}$ must
be equal to the flux of mass through the cloud surface.  Assuming spherical symmetry this surface is $4 \pi R^2$ leading to the relation
\begin{eqnarray}
\dot{M} = 4 \pi R^2 v_{\rm in} \rho_{{\rm ISM}},
\label{mass_flux}
\end{eqnarray}
where $\rho_{\rm ISM}$ is the typical ISM  density outside the cloud.
{ Note that the assumption of spherical symmetry gives an upper limit for $v_{\rm in}$ since it minimizes the surface. Consequently, our estimate of $\dot{E}_{\rm in}$ is also an upper limit. 
With Eq.~(\ref{mass_flux}) we obtain 
\begin{eqnarray}
\nonumber
 v_{\rm in} &\approx&  15 \, {\rm km \; s^{-1}}  \nonumber \\
 && \times \left( { \dot{\rm M} \over 10^{-2} \,{\rm M}_\odot {\rm yr}^{-1}} \right) \left( {n_{{\rm ISM}} \over 1 \,{\rm cm}^{-3}} \right)^{-1}
\left( { L_{\rm d} \over 80 \,{\rm pc}} \right)^{-2},
\label{vin}
\end{eqnarray}
where $L_{\rm d}$ is scaled to the diameter of a molecular cloud with mass $\sim 2 \times 10^5\,$M$_\odot$ and where $n_{\rm ISM}$ is the number density of the atomic gas using a mean molecular weight of 1.3. Thus, it appears that flows with velocities of about $15\,$km$\,$s$^{-1}$ are sufficient to explain the observed accretion rate of about 10$^{-2}\,$M$_\odot$ s$^{-1}$.

Combining Eq.'s~(\ref{eqn:dissip}) and (\ref{eqn:dissip1}) we obtain  
\begin{eqnarray}
\epsilon = {\dot{E}_{\rm decay}\over \dot{E}_{\rm in} } &\approx& {\pi^2 \rho_{\rm ISM}^2 M L^3 \sigma^3}\,{\dot{M}^{-3}}\,,
\end{eqnarray}
which translates into 
\begin{eqnarray}
\epsilon &\approx& 0.04  
 \left( { n _{\rm ISM} \over 1\,{\rm cm}^{-3} } \right)^{2}
\! \left( { M  \over 10^5 \,{\rm M}_\odot} \right) ^{\frac{6.8}{2.3}}
\! \left( { \dot{M}  \over 10^{-2} \,{\rm M}_\odot {\rm  yr}^{-1}} \right)^{-3}\,,
\label{eff_mc}
\end{eqnarray}
with the help of Eq.'s~(\ref{sigma}) and (\ref{eqn:larson2}). 

The kinetic energy associated with formation and subsequent growth of molecular clouds is sufficient to drive their internal turbulence provided that the efficiency of this conversion is not smaller than a few per cent. These numbers are in very good agreement with the numerical results discussed in Section \ref{subsec:input-rate} with mean density contrasts between cloud and intercloud medium  of several $10$s to $100$. Taking into account the higher accretion rate quoted by \cite{Fukui:2009p13386}  allows for even small efficiencies or equivalently for larger density contrasts.  
}

\section{Turbulence in Accretion Disks}
\label{sec:disks}

Last we investigate whether accretion onto T~Tauri disks may represent 
a significant contribution to the turbulence in these objects. 
{
We focus our discussion on the late stages of protostellar disk evolution, the class 2 and 3 phase, where the original protostellar core is almost completely accreted onto the central star, which is then surrounded by a remnant disk carrying only a few per cent of the total mass. Because the system is no longer deeply embedded, it is accessible to high-precision observations and structure and kinematics are well constrained \citep{Andre:2000p14204}. }
Although it is currently thought that turbulence in protostellar accretion disks is driven through 
the non-linear evolution of the magneto-rotational instability \citep{Balbus:1998p5199}, 
we believe that it is nevertheless worth to estimate the level of turbulence that the 
forcing due to accretion may sustain. 

\subsection{Are T~Tauri Disks Accreting ?}

The accretion of gas onto the disk is not easy to measure during the T Tauri phase and 
no observational data are available in the literature. On the other hand, 
the accretion from the disk onto the star has been measured in a variety of objects. Typical accretion 
rates are of the order of $3 \times 10^{-8}$ M$_\odot$ yr$^{-1}$ 
for a 1 solar mass star \cite[e.g.][]{Natta:2004p12244,Muzerolle:2005p13775,GarciaLopez:2006p12159,Gatti:2006p12193,Gatti:2008p12187}
while the mass of the disk is typically of the order of $10^{-2}$ M$_\odot$. 
Dividing the latter by the former, we find that the disk could not last more than $3 \times 10^5$ yr
which appears to be shorter than the typical T~Tauri ages {  \cite[e.g.][]{Evans:2009p15511}.}
Thus, it seems likely that
T Tauri disks are still accreting gas at a rate comparable to the one at which gas from the disk
 is accreted onto the central star \cite[see, e.g.][]{Padoan:2005p13548, Dullemond:2006p12219,Throop:2008p13748}. { Alternatively, the disk could be more massive than usually assumed \citep{Hartmann:2006p15680}.}

Another interesting observation is that the accretion rate onto the star, $\dot{M}$, is typically related to the stellar mass as 
\begin{eqnarray}
\dot{M} \propto M_\star^\alpha\,,
\label{eqn:M-Mstar}
\end{eqnarray}
where $\alpha \approx 1.8$, however, with a large scatter of about one order of magnitude \citep{Muzerolle:2003p13785,Natta:2006p12239}.  \citet{Padoan:2005p13548} and \citet{Throop:2008p13748} propose that the stars as they move through the cloud, accumulate gas at the Bondi-Hoyle accretion rate \citep{Bondi:1944p13804}. Estimating the velocity of the star to be about 1-2 km s$^{-1}$ 
and taking a mean cloud density of 10$^{3-4}$ cm$^{-3}$, they infer typical accretion onto a one solar mass star of about $10^{-8}$ M$_\odot\,$yr$^{-1}$. In particular, to get accretion rates compatible with the largest observed  values, they need to invoke large densities of $10^{4-5}$ cm$^{-3}$.
{  However, the filling factor of gas at such densities is very low, of order of a few per cent. Densities  of $10^{4-5}\,$cm$^{-3}$ typically correspond to cores with sizes of $\sim 0.1\,$pc. If we take typical stellar velocities of $1-2\,$km$\,$s$^{-1}$, it requires only 10$^5\ $yr to cross a core. Most of the time T~Tauri stars travel through low density gas,  eventually building up a widely dispersed population \cite[see, e.g.][]{Neuhaeuser:1995p17478, Wichmann:1996p17489}
}

Here we propose that accretion { during the  late phases of protostellar evolution} proceeds in a different way. { We base our discussion on the assumption that the velocity of the star is inherited from the bulk velocity of the core in which it forms.} {  
In $\rho$-Oph, for example, the typical core-to-core velocity dispersion is found to be less than $0.4\,$km$\,$s$^{-1}$ \cite[][]{Andre:2007p25886}. The typical stellar velocity dispersion measured in nearby  T~Tauri associations or open star clusters is of similar order, $\sim0.3\,$km$\,$s$^{-1}$ for the Hyades \citep{Madsen:2003p26445}, $\sim0.5 - 0.6\,$km$\,$s$^{-1}$ for Coma Berenices, Pleiades, and Praesepe \citep{Madsen:2002p26491}, and below $\sim1\,$km$\,$s$^{-1}$ for $\alpha$ Per \citep{Makarov:2006p26414}, Lupus \citep{Makarov:2007p26369}, and the sub-groups in Taurus \citep{Jones:1979p26561,Bertout:2006p25682}. It gets above $1\,$km$\,$s$^{-1}$ only for the more massive clusters and OB associations \citep{Madsen:2002p26491}.}
Since {  star forming cores are part} of a turbulent molecular cloud, {  we resort once again to Larson's relation and assume} the mean velocity of any fluid element with respect to the center of the core is increasing with distance from the core as $\sim 0.46 \times (L/1\,{\rm pc})^{0.5}$. This expression is identical to Eq.~(\ref{sigma}) except for the factor $3^{-1/2}$ which comes from the fact that we consider $\sigma_{\rm 1D}$ instead of $\sigma$.
It takes a star about $2.2 \times 10^6\ $yr to reach a distance of 1$\,$pc if it travels with 0.46$\,$km$\,$s$^{-1}$. This  implies that during a long period of time, comparable to the age of the T~Tauri star, its velocity with respect to the surrounding gas is not of the order of $1-2\,$km$\,$s$^{-1}$ but more comparable to the sound speed of the gas $c_{\rm s} \approx 0.2\,$km$\,$s$^{-1}$.
As we show below, we can {quantitatively} reproduce the observational 
relation $\dot{M} \simeq 2-3 \times 10^{-8} \; {\rm M}_\odot \,{\rm yr}^{-1} \; (M_\star/1\,{\rm M}_\odot)^{1.8}$ obtained by \citet{Natta:2006p12239} if we assume that the accretion onto the star is essentially controlled by the accretion onto the disk from the turbulent cloud environment. { In this picture, the disk only acts as buffer for this overall accretion flow \cite[possibly leading to occasional outbursts, see][]{Hartmann:1996p15524}.}

{ 
Our estimate is very close to the spherical accretion considered by \citet{Bondi:1952p13808}. The difference however
is that  turbulence velocity increases with the distance \cite[see also][for accretion in fractal media]{Roy:2007p18500}.
Consider a star of mass $M_\star$ inside a turbulent cloud. Assuming that the star is at rest, it will be able to accrete 
gas inside a sphere of radius $R_{\rm acc}$, such that fluid particles inside this radius are 
gravitationally bound to the star,
\begin{eqnarray}
R_{\rm acc} \approx {GM_\star \over c_{\rm s}^2 + \sigma_{1D} ^2}\,,
\end{eqnarray}
where $c_{\rm s} \approx 0.2\,$km$\,$s$^{-1}$ is the sound speed and $\sigma_{1D} = 3^{-1/2} \sigma$ is the 1-dimensional velocity dispersion, which we obtain from Larson's relation, Eq.~(\ref{sigma}), assuming isotropy.
From this relation we can easily obtain $R_{\rm acc}$ as a function of mass
\begin{eqnarray}
R_{\rm acc}={3 L_0 \over 2 v_0^2 } \left(  - c_{\rm s}^2 + \sqrt{c_{\rm s}^4 + {4 M_\star G v_0^2 \over 3 L_0} } \,\right),
\end{eqnarray}
where $v_0 \approx 0.8\,$km$\,$s$^{-1}$ and $L_0=1\,$pc.
The total mass included into the sphere of radius $R_{acc}$ is simply $M_{\rm acc}=4 \pi /3 R_{\rm acc}^3 \,\bar{\rho} $
where $\bar{\rho}$ is the mean density inside molecular clouds, which we take to be of the order of
 $\mu \times $100 cm$^{-3}$ with $\mu=2.35 \times m_{\rm H}$ being the mean molecular weight and $m_{\rm H}$ being the mass of the hydrogen atom.
This mass falls into the star/disk system on a timescale of the order of $\tau_{\rm acc} \approx (G M_\star / R_{\rm acc}^3)^{-1/2}$
leading to a typical accretion rate of
\begin{eqnarray}
\dot{M}_{\rm acc} \approx { (4 \pi /3) R_{\rm acc}^3 \bar{\rho} \over \tau _{\rm acc} } = (4 \pi /3) R_{\rm acc}^{3/2} M_\star^{1/2} G^{1/2} \bar{\rho}. 
\label{accret_form}
\end{eqnarray}
Note that as the matter inside $R_{\rm acc}$ is accreted, the pressure drops and it is replenished in 
about a crossing time by the surrounding gas. Since by definition of $R_{\rm acc}$, this crossing time is comparable to the accretion time, there is always fresh gas available for accretion. 
Altogether, we obtain
\begin{eqnarray}
\dot{M}_{\rm acc} \propto  M_\star^{1/2} \left(  -1 + \sqrt{1+\xi M_\star}
\right)^{3/2}\,,
\end{eqnarray}
with $\xi = 4Gv_0^2/(3L_0 c_{\rm s}^4)$.

\begin{figure}[tbp]
\centerline{\psfig{file=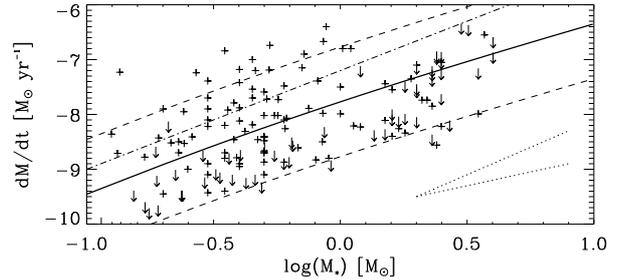,width=8.5cm}}
\caption{Prediction of the accretion rate onto the disk as a function of the mass of the star. The solid line corresponds to a mean density of $\bar{n}=100\,$cm$^{-3}$ while the two dashed lines are for $\bar{n}=1000\ $cm$^{-3}$ (upper curve) and $\bar{n}=10\ $cm$^{-3}$ (lower curve).  To guide your eye the dotted lines indicate the slope of the relations $\dot{M} \propto M_\star^2$ and $\dot{M} \propto M_\star$. We compare with data from \citet{Calvet:2004p14672}, \citet{Mohanty:2005p14921}, \citet{Muzerolle:2005p13775}, and \citet{Natta:2006p12239} as displayed in Figure 3 of \citet{GarciaLopez:2006p12159},  where crosses indicate detections and arrows upper limits. The dot-dashed line is the fit proposed by  \citet{Natta:2006p12239} with slope $\alpha = 1.8$.}
\label{accret_disk}
\end{figure}

In the limit where the cloud is not turbulent ($\sigma^2 \ll c_{\rm s}^2$, i.e.\ $\xi \ll 1$) this leads to 
$\dot{M} \propto M_\star^2$, while if it is dominated by turbulence  ($\sigma^2 \gg c_{\rm s}^2$ or $\xi \gg 1$) we obtain $\dot{M} \propto M_\star^{5/4}$. The slope of the $\dot{M}_{\rm acc}\,$-$\,M_\star$ relation therefore lies between 1.25 and 2. For a 1 solar mass star, $R_{\rm acc} \approx 0.1\,$pc and $\sigma \approx 0.25\,$km$\,$s$^{-1} \approx c_{\rm s}$, leading to a slope of $\sim 1.8$.  Figure~\ref{accret_disk} shows the dependency of $\dot{M}$ as specified by Eq.~(\ref{accret_form}) (solid line).}
As molecular clouds are highly inhomogeneous the local density can vary over orders of magnitude. To account for this variation, we also display the values of the accretion rate as predicted by Eq.~(\ref{accret_form}) for $\bar{n}=10$ and 1000 cm$^{-3}$ (dashed lines). To connect to real measurements, we overplot the data presented by \citet{GarciaLopez:2006p12159}. The agreement is remarkable. Despite its simplicity, our model provides a good order-of-magnitude fit. In particular, also our prediction that for masses between 1 and 10$\,$M$_\odot$ the relation between accretion rate and stellar mass becomes more shallow seems to be confirmed by the observational data.  

Finally, we  stress that accretion of the type we consider here, is difficult to avoid as long as the star remains within its parent molecular cloud and we conclude that continuous accretion onto the disk even during the T~Tauri phase is an interesting phenomenon that deserves further attention.

\subsection{Expected Velocity Dispersion in Accretion Disks}

{ 
Considering Eq.'s.~(\ref{eqn:dissip}) and (\ref{eqn:dissip1}), we again take the typical turbulent length scale to be comparable to the disk thickness, $L_{\rm d} = 2 H$, where $H$ is the vertical scale height. Since the gravitational potential is dominated by the central star, we estimate the infall velocity $v_{\rm in}$ to be 
\begin{eqnarray}
v_{\rm in} = \left(2 G M_\star \over R\right)^{1/2}\,,
\label{Vgrav}
\end{eqnarray}
and get
\begin{eqnarray}
\label{eff_disk}
 \epsilon  = { \dot{E}_{\rm decay} \over  \dot{E}_{\rm in}} &=& \frac{M_{\rm disk} \sigma^3/2H}{\dot{M}v_{\rm in}^2} \\
 & \approx& \frac{3^{3/2}}{2} {\cal M}^3 \left(\frac{\dot{M}}{c_{\rm s}^3/G}\right)^{-1} \left(\frac{M_{\rm disk}}{M_\star}\right) \left(\frac{H}{R}\right)^{-1} \,, \nonumber
\end{eqnarray}
where $M_{*}$ and $M_{\rm disk}$ are the masses of the central star and its disk, respectively, and where ${\cal M} = \sigma_{\rm 1D} / c_{\rm s}$ is the rms Mach number of the turbulence in the disk.

For a temperature of $10\,$K, $c_{\rm s} \approx 0.2\,$km$\,$s$^{-1}$ and the collapse rate $c_{\rm s}^3/G \approx 2 \times 10^{-6}\,$M$_\odot\,$yr$^{-1}$. With the fiducial value $\dot{M} \approx 2  \times 10^{-8}\,$M$_\odot\,$yr$^{-1}$ for a solar mass T~Tauri star, the expression in the first bracket in Eq.~(\ref{eff_disk}) is roughly 100. In the class 2 and 3 phase, the ratio $M_{\rm disk}/M_\star \approx 0.01$. We can estimate the ratio of local scale height and radius using Appendix \ref{app:potential-method} which gives $H/R \approx 0.1$ for $M = 1\,$M$_{\odot}$ and a disk radius  of $200\,$AU. Together, this leads to 
\begin{eqnarray}
 \epsilon &\approx& 24\,  {\cal M}^3  \,.
\end{eqnarray}
Because ${\cal M} \approx \left( \epsilon / 24 \right)^{1/3}$, late accretion  can only drive subsonic turbulence in T~Tauri disks. If the true efficiency of the process is of the order of 0.1, then the maximum rms Mach number that can be reached is ${\cal M} \approx 0.16$. If $\epsilon = 0.01$ then ${\cal M} \approx 0.07$. Only few 
%
%
%
%
%
%
measurements of  the velocity dispersion in disks have been carried out to date. \cite{Dutrey:2007p13551} report values of $\sim 0.1\,$km$\,$s$^{-1}$  which corresponds to a Mach number ${\cal M} \approx 0.5$. Although this value is slightly larger (factor 3 to 7) than the ones we inferred, there are large theoretical and observational uncertainties. We have not taken into account that accretion, particularly if it is non-axisymmetric may help to tap energy from the rotational energy available in the disk. We  conclude that our simple estimates support the hypothesis that accretion can provide at least some level of turbulence in circumstellar T~Tauri disks. Note that even if it can sustain a  significant amount of turbulence, the question as to whether this turbulence could transport angular momentum efficiently is entirely open.}

Because $\dot{M}$ varies with $M_\star$ as illustrated in Figure \ref{accret_disk}, the ratio $\dot{E}_{\rm decay} /  \dot{E}_{\rm in}$ depends on  $M_\star$ as well. According to Eq.~(\ref{eff_disk}),  the rms Mach number of the accretion-driven turbulence for any given efficiency is smaller in disks around low-mass stars than around high-mass stars.  For a low-mass star with $M=0.1\,$M$_{\odot}$ an efficiency of 10\%  would correspond to ${\cal M} =0.02$ only, while for $M=10\,$M$_{\odot}$ $\epsilon = 0.1$ would lead to ${\cal M} = 0.2$, assuming the standard values $H/R = 0.1$ and $M_{\rm disk}/M_{\star} = 0.01$ in both cases.

%

\section{Conclusion}
\label{sec:end}

When cosmic structures form, they grow in mass via accretion from their surrounding environment. This transport of material is associated with kinetic energy and provides a ubiquitous source of driving internal turbulence. In this paper we propose that the turbulence that is ubiquitously observed in astrophysical objects on all scales at least to some degree is driven by this accretion process. To support our idea, we combined analytical arguments and results from numerical simulations of converging flows to estimate the level of turbulence that is provided,  and applied this theory to galaxies, molecular clouds, and protostellar disks.  

We first studied the Milky Way as well as 11 galaxies from the THINGS survey, and found that in Milky Way type galaxies the level of turbulence ubiquitously observed in the atomic gas in the disk can be explained by accretion, provided the galaxies accrete gas at a rate comparable to the rate at which they form stars. { Typically, the efficiency required to convert infall motion into turbulence is of order of a few percent. This process is particularly relevant in the extended outer disks beyond the star-forming radius where stellar sources cannot provide alternative means of driving turbulence.  It is attractive to speculate that the population of high-velocity clouds, e.g.\ as observed around the Milky Way, is the visible signpost for high-density peaks in this accretion flow.  The assumption of steady state evolution, however, fails for dwarf galaxies. In order to drive the observed level of turbulence the accretion rate needs to exceed the star formation rate by far and we expect other sources to dominate. } We also applied our theory to the dense star-forming knots in chain galaxies at high redshift and, in agreement with previous studies, came to the conclusion that their turbulence could be driven by accretion as well.  

We then turned to molecular clouds. Using the recent estimate by Fukui et al. (2009) of the accretion rate within molecular clouds in the large Magenallic cloud, we found that  accretion is sufficient to drive their internal turbulence at the observed level. This is in good agreement with the finding that most of the turbulent energy  in molecular clouds carried by large-scale modes \cite[see, e.g.][]{Ossenkopf:2002p334,Brunt:2003p17014,Brunt:2009p16966} and also with the fact that clouds which do not form massive star show the same amount of turbulence as those which do. This excludes internal sources. It is the very process of cloud formation that drives turbulent motions on small scales by establishing the turbulent cascade. Numerical simulations of colliding flows reveal that the turbulence within dense clumps generated by converging flows of incoming speed of $15-20\,$km s$^{-1}$ is fully compatible with Larson's relations.

As no observational evidence for accretion onto T~Tauri disks has been reported, our investigation of accretion driven turbulence in protostellar disks is more speculative. However, very similar to galactic disks, without late mass accretion protostellar disks would drain onto their central stars on timescales shorter than the inferred disk lifetimes. Using this as starting point, we were able to show that disk accretion can drive subsonic turbulence in T~Tauri disks at roughly the right level if the rate at which gas falls onto the disk is comparable to the rate at which disk material accretes onto the central star. This process also provides a simple explanation for the observed relation of accretion rate and stellar mass,   $\dot{M} \propto M_\star^{1.8}$.

We conclude that accretion-driven turbulence is a universal concept with far-reaching implications for a wide range of astrophysical objects.

\acknowledgement
We thank Edvige Corbelli, Francesco Palla, and Filippo Mannucci for organizing an excellent and highly interesting conference on the Schmidt-Kennicutt relation in Spineto, which triggered this work. We are grateful to Javier Ballersteros-Paredes, Frank Bigiel, Robi Banerjee,  Paul Clark, Kees Dullemond, Simon Glover, Adam Leroy, Mordecai Mac Low, Rahul Shetty, and Fabian Walter for many stimulating discussions, and Lee Hartmann, Eve Ostriker, Joseph Silk for valuable suggestions that helped to improve this paper.  We thank our referee Enrique V\'azquez-Semadeni for very insightful comments, and 
Domenico Tamburro for sending the processed dataset of the 11 THINGS galaxies discussed here, as well as Antonella Natta,  Elisabetta Rigliaco and Leonardo Testi for providing data of protostellar accretion rates and disk masses.
R.S.K.\ thanks for the warm hospitality of the {\em Ecole normale sup\'erieure} in Paris. R.S.K.\ acknowledges financial support from the German {\em Bundesministerium f\"{u}r Bildung und Forschung} via the ASTRONET project STAR FORMAT (grant 05A09VHA) and from the {\em Deutsche Forschungsgemeinschaft} (DFG) under grants no.\ KL 1358/1, KL 1358/4, KL 1359/5, KL 1358/10, and KL 1358/11. R.S.K.\ furthermore thanks for subsidies from a Frontier grant of Heidelberg University sponsored by the German Excellence Initiative and for support from the {\em Landesstiftung Baden-W{\"u}rttemberg} via their program International Collaboration II (grant P-LS-SPII/18).
This work was granted access to the HPC resources
of  CINES under the allocation  x2009042036
made by GENCI (Grand Equipement National de Calcul Intensif).

\appendix

{
\section{Motivation of the Potential Method}
\label{app:potential-method}
The gravitational potential in rotationally supported disks is often approximately spherical symmetric. This holds for the extended outer HI disks in dark matter dominated spiral galaxies (Section  \ref{sec:galaxies}) as well as for protostellar disks in the late phase of the evolution where most of the system mass is carried by the central star (Section  \ref{sec:disks}). The error is usually less than a few per cent. The equation of hydrostatic balance reads
\begin{eqnarray}
\vec \nabla P &\approx& - \vec F_{\rm g},
\end{eqnarray}
with pressure $P = c_{\rm s}^2 \rho$ and gravitational force  $F_{\rm g}  = GM \rho /\left( R^2 + z^2\right)$ pointing towards the center. We use cylindrical coordinates, and  $c_s$ and $M$ are sound speed and enclosed mass, respectively. The pressure gradient in vertical direction is \begin{eqnarray}
\frac{dP}{dz} &\approx& \frac{\rho c_{\rm s}^2}{H}\,,
\end{eqnarray}
where $H$ is the vertical scale height. The $z$ component of the force at any location $H\ll R$ is approximately  
\begin{eqnarray}
F_{{\rm g},z}&\approx& \frac{GM\rho}{R^2 + H^2} \sin\left(\frac{H}{\sqrt{R^2+H^2}} \right)\ \approx \frac{GM\rho}{R^2}\frac{H}{R}\,,
\end{eqnarray}
and consequently
\begin{eqnarray}
\frac{H}{R} &\approx& \frac{c_{\rm s}}{\sqrt{GM/R}} \approx \frac{c_{\rm s}}{v_{\rm rot}},
\end{eqnarray}
with the circular velocity $v_{\rm rot} = \left(GM/R\right)^{1/2}$. This is Eq.~(\ref{eqn:total-pot}).
}

{
\section{The $M\,$-$\,\sigma^3$ Relation at High Densities}
\label{app:M-sigma}
In this appendix we discuss the validity of the relation $\epsilon \propto
\rho^{-1}$ inferred in Section \ref{sec:concept}.

First, we note from Fig.~\ref{efficiency} that at high densities
the velocity dispersion of the dense gas remains nearly constant 
when the density threshold varies. This is clearly due to the fact that 
the total mass of the dense gas is constituted by several dense clumps 
which are not spatially correlated and randomly distributed in the 
turbulent box (as can be seen from Fig.~\ref{col_dens}). Thus, 
the velocity dispersion simply reflects the velocity dispersion of the box 
which typically varies with distance $l$ as $l^{1/3-1/2}$.
Therefore, in this regime of density the integral of interest
$\int \rho v^3 dV$ can be approximated as $\sigma^3 \int \rho  dV$
where $\sigma$ is approximately the velocity dispersion 
corresponding to the size of the more distant clumps. 
As a consequence, the dependence of $M \sigma^3$ on the density threshold
 should in this regime be identical to the one of the mass. 

For simplicity, let us consider isothermal gas. In this case, the density PDF is approximately log-normal  \cite[][]{VazquezSemadeni:1994p13304,Padoan:1997p18754,Klessen:2000p236,Kritsuk:2007p14285}, however,  with higher-order corrections that become significant for highly compressive forcing \cite[][]{Federrath:2008p818,Federrath:2009p14029}, 
\begin{eqnarray}
P(\rho)={1 \over \sqrt{2 \pi \sigma^2}} 
\exp\left[ - {(\delta-\bar{\delta})^2 \over 2 \sigma^2 } \right]\,,
\label{pdf}
\end{eqnarray}
where $\delta=\log(\rho/\bar{\rho})$, $\bar{\delta} = - \sigma^2/2$ and $\sigma^2=\ln (1+ b^2 {\cal M}^2)$
with parameter $b \simeq 0.5$ and Alfv\'enic Mach number ${\cal M} \simeq 5$.

The mass above some density threshold $\rho_{\rm t}$ is obtained as
\begin{eqnarray}
M_{\rm t} = \bar{\rho} V {1 \over \sqrt{2 \pi \sigma^2}} 
\int ^{\infty} _{\delta_{\rm t}} \exp(\delta) \exp \left[ - {(\delta-\bar{\delta})^2 \over 2 \sigma^2 }  \right] d \delta,
\label{mass_t}
\end{eqnarray}
where $\delta_{\rm t}= \log (\rho_{\rm t}/ \bar{\rho})$ and thus
\begin{eqnarray}
f_{\rm t}=
M_{\rm t} / (\bar{\rho} V) &=&  {1 \over \sqrt{ \pi }} 
\int ^{\infty} _{ {\delta_{\rm t} - \sigma^2/2 \over \sqrt{2} \sigma } } \exp \left( - x^2  \right) dx, \\
 &=& {1 \over 2} \bar{\rho} V \left[ 1 - {\rm erf} \left( {\delta_{\rm t} - \sigma^2/2 \over \sqrt{2} \sigma } \right) \right].
\nonumber
\label{mass_int}
\end{eqnarray}

\begin{figure}[tbp]
\centerline{\psfig{file=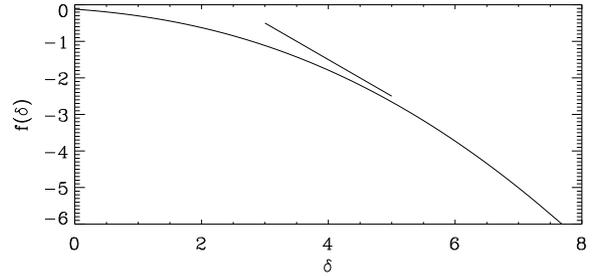,width=8.5cm}}
\caption{Fraction $f_{\rm t}$ of mass above a threshold density $\rho_{\rm t}$ as function of density logarithmic density contrast $\delta _{\rm t} $.  For reference, the straight line indicates a slope of $-1$.}
\label{M_rho_theo}
\end{figure}
 
Figure~\ref{M_rho_theo} shows $f_{\rm t}$ as function of $\delta_{\rm t}$. Its behavior is similar to the mass distribution displayed in Fig.~\ref{efficiency}
and the relation $M \sigma^3 \propto \rho^{-1}$ is approximately recovered for 
$\delta_{\rm t}$ above $\sim 4$ (where $\sigma$ is observed to be nearly constant).
Thus, keeping in mind that the simulations include cooling, gravity and 
magnetic fields, we conclude that the effect we observe is reasonably well 
reproduced by this simple approach. At higher density contrast $\delta_{\rm t}$,  the slope of $f_{\rm t}$ 
becomes steeper and we expect $\epsilon$ to decrease more rapidly.
}

\bibliographystyle{aa}
\bibliography{library}

\end{document}